\documentclass[11pt,a4paper]{article}
\usepackage{jheppub}
    \makeatletter
\def\@fpheader{\relax}
\makeatother
\usepackage{mathtools}
\usepackage{graphicx}
	\graphicspath{ {images/} }
\usepackage{blindtext}	
\usepackage{amsmath,amssymb,amsthm}
\usepackage{latexsym,graphicx,color}
\usepackage{enumerate}
\usepackage{color}

\def\half{\frac{1}{2}}
\def\a{\alpha}

\def\half{\frac{1}{2}}
\usepackage{hyperref}

\newcommand{\D}{\mathcal{D}}
\newcommand{\p}{\partial}
\newcommand{\Tr}{{\rm Tr}}

\newcommand{\bea}{\begin{eqnarray}}
\newcommand{\eea}{\end{eqnarray}}
\def\Tr{ \hbox{\rm Tr}}

\def\l{\left}
\def\r{\right}

\title{{\bf Confinement and moduli locking of Alice strings and monopoles 
}}
\author{Muneto Nitta}
 \affiliation{Department of Physics $\&$ Research and Education Center for Natural Sciences,\\ Keio University, Hiyoshi 4-1-1, Yokohama, Kanagawa 223-8521, Japan}
 \emailAdd{nitta(at)phys-h.keio.ac.jp}

\date{\today}

\abstract{
We argue that strings (vortices) and monopoles are confined,
when fields receiving nontrivial Aharonov-Bohm (AB) phases around a string 
develop vacuum expectation values (VEVs).
We illustrate this in  
an $SU(2) \times U(1)$ gauge theory with charged triplet complex 
scalar fields admitting Alice strings and monopoles, 
by introducing charged doublet scalar fields 
receiving nontrivial AB phases around the Alice string. 
The Alice string carries a half $U(1)$ magnetic flux and 
$1/4$ $SU(2)$ magnetic flux taking a value in  
two of the $SU(2)$ generators characterizing the $U(1)$ modulus.
This string is not confined in the absence of a doublet VEV in the sense that 
the $SU(2)$ magnetic flux can be detected at large distance by an AB phase 
around the string. 
When the doublet field develops VEVs, there appear two kinds of phases 
that we call deconfined and confined phases.
When a single Alice string is present 
 in the deconfined phase,  
the $U(1)$ modulus of the string and the vacuum moduli are locked
(the bulk-soliton moduli locking). 
In the confined phase, 
the Alice string is inevitably attached by a domain wall that we call an AB defect 
and is confined with an anti-Alice string or 
another Alice string with the same $SU(2)$ flux. 
Depending on the partner, the pair annihilates 
or forms a stable doubly-wound Alice string 
having an $SU(2)$  magnetic flux inside the core, 
whose color cannot be detected at large distance by AB phases, 
implying the ``color'' confinement.
The theory also admits stable 
Abrikosov-Nielsen-Olesen string and a ${\mathbb Z}_2$ string 
in the absence of the doublet VEVs, and  
each decays into two Alice strings 
in the presence of the doublet VEVs.
A monopole in this theory can be constructed as a closed Alice string 
with the $U(1)$ modulus twisted once, 
and we show that with the doublet VEVs, monopoles are also 
confined to monopole mesons 
of the monopole charge two.
}
\begin{document}

\maketitle

\section{Introduction}
Color confinement is one of the most challenging problems 
in modern high energy physics. 
Particles transforming under color gauge group cannot be observed 
and are confined. 
Here, we discuss 
Aharonov-Bohm (AB) effects  \cite{Aharonov:1959fk} 
may be one of key ingredients to understand color confinement. 
When a charged particle scatters from 
a solenoid with non-zero magnetic flux inside,
a gauge potential rather than a field strength (a magnetic or electric field) 
yields an AB phase to the charged particle
\cite{Aharonov:1959fk},
resulting in a non-trivial differential scattering cross section. 
Such an AB effect was experimentally observed
\cite{Tonomura:1982,Tonomura:1986}. 
In the theory side, cosmic strings (vortices) exhibiting AB effects were proposed in cosmology \cite{Alford:1988sj,Vilenkin:1991zk,MarchRussell:1991az}  
as well as in string theory 
\cite{Polchinski:2005bg,Ookouchi:2013gwa,
Harvey:2007ab,Harvey:2008zz,Okada:2014wma}.  
Non-Abelian vortices in supersymmetric gauge theories 
\cite{Hanany:2003hp,Auzzi:2003fs,
Eto:2004rz,Eto:2005yh,Eto:2006cx,
Tong:2005un,Eto:2006pg,Shifman:2007ce,Shifman:2009zz}
exhibit AB effects once a part of flavor symmetry is gauged \cite{Evslin:2013wka,Bolognesi:2015mpa,Bolognesi:2015ida}. 
In dense QCD, 
a color magnetic flux tube in the 2SC phase exhibits AB effects for quarks 
\cite{Alford:2010qf}, and  
 a non-Abelian vortex (color magnetic flux tube) in the color-flavor locked phase 
 \cite{Balachandran:2005ev,Nakano:2007dr,Eto:2009kg,Eto:2009bh,Eto:2009tr,Eto:2013hoa} 
 exhibits (electromagnetic) AB effects for charged particles
\cite{Chatterjee:2015lbf} 
as well as ${\mathbb Z}_3$ (color) AB effects for quarks  
 \cite{Cherman:2018jir,Chatterjee:2018nxe,Chatterjee:2019tbz,Hirono:2018fjr,Hirono:2019oup}.
 Non-Abelian Alice strings in two-flavor dense QCD 
 exhibit nontrivial AB phases \cite{Fujimoto:2020dsa}.

Alice strings are not only 
an example of AB strings but also exhibit a peculiar electromagnetic property.
 When electrically charged particles encircle around an Alice string, 
the signs of their electric charges are flipped 
because of non-single valuedness of an electromagnetic  generator 
\cite{Schwarz:1982ec, Kiskis:1978ed}.
The simplest Alice string is present in  
an $SO(3)$ gauge theory with 
fiveplet scalar fields (real traceless symmetric tensor), 
where the $SO(3)$ gauge symmetry is spontaneously broken down to $O(2)$ identified with the electromagnetic $U(1)$ group, 
which is not singly defined around the string 
 \cite{Schwarz:1982ec, Striet:2000bf,Bais:2002ae,Striet:2003na,Benson:2004ue}. 
In quantum field theory, Alice strings were extensively studied  
for their exotic properties such  
 as topological obstruction, non-local (Cheshire) charge, and non-Abelian statistics 
 \cite{Alford:1990mk, Alford:1990ur, Preskill:1990bm,  Alford:1992yx, Bucher:1992bd, Bucher:1993jj, Lo:1993hp,Striet:2000bf,Benson:2004ue}. 
 In condensed matter physics, a global analogue of Alice strings was found 
 in spinor Bose-Einstein condensates of ultracold atomic gases  
 \cite{Leonhardt:2000km,Ruostekoski:2003qx,Kobayashi:2011xb,Kawaguchi:2012ii}. 
Recently, 
  a $U(1) \times SU(2)$ (supersymmetric) 
  gauge theory coupled with complex triplet scalar fields    
was found to admit 
  a Bogomol'nyi-Prasad-Sommerfield (BPS) \cite{Bogomolny:1975de,Prasad:1975kr} 
  Alice string  \cite{Chatterjee:2017jsi,Chatterjee:2017hya,Chatterjee:2019zwx}.\footnote{
  Actually, this model is a local (gauged) version of the theory admitting global Alice strings in Refs.~\cite{Leonhardt:2000km,Ruostekoski:2003qx,Kobayashi:2011xb,Kawaguchi:2012ii}
  in which the symmetry group 
  $U(1) \times SU(2)$ is fully global. 
Recently, a hybrid of local and global Alice string was 
studied in the  axion cosmology 
\cite{Sato:2018nqy,Chatterjee:2019rch}, 
in which the $U(1)$ part is global and the $SU(2)$ part is local.   
The opposite case in which 
the $U(1)$ part is local and the $SU(2)$ part is global 
is also known in certain superconductors \cite{Materne2015}.
} 
The theory also admits 
a stable (BPS) Abrikosov-Nielsen-Olesen (ANO) string carrying a unit $U(1)$ magnetic flux
\cite{Abrikosov:1956sx,Nielsen:1973cs}, 
and a ${\mathbb Z}_2$ string 
connecting the two elements of the center of $SU(2)$
and consequently carrying one half $SU(2)$ magnetic flux
\cite{Nielsen:1973cs}.\footnote{
This is one half compared with the one generated by a closed loop in $SU(2)$.
}
In this case, the Alice string is a hybrid of these two kinds of vortices 
and carries one half $U(1)$ magnetic flux of an ANO vortex 
and one quarter $SU(2)$ magnetic flux
 taking a value in  
two of the $SU(2)$ generators characterizing the $U(1)$ modulus.
On the other hand, 
a monopole in this theory can be constructed as a closed Alice string 
with the $U(1)$ modulus twisted once, 
which was found in the global analogue \cite{Ruostekoski:2003qx}.
Such a construction of monopole was also known in the conventional Alice theory, the $SO(3)$ gauge theory with fiveplet scalars  \cite{Striet:2000bf,Bais:2002ae,Striet:2003na,Benson:2004ue}. 

In the previous paper, 
in order to discuss what happens when a field receiving a non-trivial AB 
phase develops a vacuum expectation value (VEV),
we further introduced  doublet scalar fields 
receiving non-trivial AB phases in the presence of an Alice string, 
and found that a soliton or domain wall which we call an AB defect\footnote{
Another example of an AB defect can be found in 
the Georgi-Machacek model \cite{Georgi:1985nv} 
proposed as a model beyond the Standard Model (SM), 
consisting of three real triplet scalar fields and one doublet scalar field.
If the triplet VEVs are larger than the doublet VEV,
then a ${\mathbb Z}_2$ string is attached by an AB defect \cite{Chatterjee:2018znk}.
}
is attached to the Alice string.\footnote{
Although a string attached by an AB defect looks similar to 
a more conventional 
configuration of a string attached by a domain wall
\cite{Kibble:1982dd,Vilenkin:1982ks},
we emphasize that the mechanisms are rather different.
For the conventional case, an explicit symmetry breaking term 
of the $U(1)$ symmetry around which a string winds 
induces a domain wall. 
String-wall composites in axion models 
\cite{Kawasaki:2013ae},
two Higgs doublet models  
\cite{Eto:2018hhg,Eto:2018tnk},
and domain wall skyrmions 
\cite{Nitta:2012xq,Kobayashi:2013ju}
belong to this class. 
We may classify the two cases as follows; 
The conventional case is induced by  
``explicit symmetry breakings,'' while the AB defect is induced by 
``spontaneous symmetry breakings.''
}
This theory admits multiple BPS Alice strings at arbitrary positions, when we turn off the doublet field. Since a doublet encircling two strings receives no AB phase,  the two Alice strings should be connected by one AB defect when the doublet develops a VEV. One natural question was how each Alice string find a partner, when there are many Alice strings.

In this paper, we define the ``color'' (de)confinement as a situation that a color 
cannot (can)  be detected 
at large distance by AB phases of some fields belonging to 
a representation of the color group.
We then argue that the Alice string is not confined in the absence of a doublet VEV 
because
the $SU(2)$ (color) magnetic flux can be detected at large distance 
by an AB phase of the doublet fields around the string, 
while the confinement occurs
once the doublet field develops VEVs (in the certain phase defined below).\footnote{
In this definition, although 
 a non-Abelian vortex 
   in the CFL phase of dense QCD
 carries a color magnetic flux 
 \cite{Balachandran:2005ev,Nakano:2007dr,Eto:2009kg,Eto:2009bh,Eto:2009tr,Eto:2013hoa}, 
 it exhibits only a ${\mathbb Z}_3$ AB phase within $SU(3)$ color symmetry, which is {\it color singlet}.
Namely, a color flux of the non-Abelian vortex cannot be seen from large distance,  
and consequently it is already confined as it is. 
The same holds for non-Abelian vortices in supersymmetric gauge theories 
\cite{Hanany:2003hp,Auzzi:2003fs,
Eto:2004rz,
Eto:2005yh,Eto:2006cx,
Tong:2005un,Eto:2006pg,Shifman:2007ce,Shifman:2009zz}
unless 
a part of flavor symmetry is gauged \cite{Evslin:2013wka,Bolognesi:2015mpa,Bolognesi:2015ida}.
}
When the doublet field develops VEVs, 
there appear deconfined and confined phases with unbroken  
${\mathbb Z}_4$ and ${\mathbb Z}_2$ symmetries, respectively.
In the confined phase, 
the Alice string is inevitably attached by an AB defect 
and is confined with an anti-Alice string or 
another Alice string with the same $SU(2)$ flux. 
Depending on the partner, the pair annihilates 
or forms a stable doubly-wound Alice string 
having an $SU(2)$  magnetic flux inside the core, 
where the ``color'' cannot be seen at large distance by AB phases, 
implying the ``color'' confinement.
We also show that with the doublet VEVs, monopoles are also 
confined to pair annihilate or to form monopole mesons 
of the monopole charge two.
We thus call the unbroken phase without the doublet VEVs 
and the confined phase with the doublet VEVs as 
deconfined and confined phases, respectively. 

Another interesting phenomenon is 
that in the presence of a single Alice string, 
the phase is enforced to be in the deconfined phase 
to annihilate an AB defect, 
and consequently the $U(1)$ modulus of the string and the vacuum moduli of the doublet fields 
are locked.
We call this phenomenon as the bulk-soliton moduli locking. 
We also show that while  
an ANO string 
and a ${\mathbb Z}_2$ string 
are stable in the absence of the doublet VEVs,
each decays into two Alice strings to each of which an AB defect is attached, once the doublet field develops the VEVs.
We  discuss that whether a set of two strings exhibits confinement or decay can be determined from AB phases of the doublet field encircling the two strings.

This paper is organized as follows.
In Sec.~\ref{sec:model}, we introduce our model admitting 
BPS Alice strings and discuss in detail 
symmetry breaking patterns  
when the triplet field develops VEVs at high energy 
and the doublet field also develops VEVs at low energy.
We also discuss the opposite hierarchy 
in which 
the doublet field develops VEVs at high energy 
and the triplet field develops VEVs at low energy.
In Sec.~\ref{sec:string-monopole}, we introduce 
an ANO string, a ${\mathbb Z}_2$ string, an Alice string, 
and an Alice monopole. 
In Sec.~\ref{sec:confinement}, 
we first discuss 
AB phases of the doublet fields around a single Alice string 
and around two strings.
We then discuss, in the presence of doublet VEVs, 
the bulk-soliton moduli locking, 
confinement of Alice strings 
and monopoles, 
and decay of ANO and ${\mathbb Z}_2$ strings.
Sec.~\ref{sec:summary} is devoted to a summary and discussion.
In Appendix \ref{sec:app0}, we show that complex adjoint scalar fields 
are equivalent to complex $2 \times 2$ symmetric tensor 
scalar fields.
In Appendix \ref{sec:app}, we give a brief review of previous results for BPS Alice string 
solution 
\cite{Chatterjee:2017jsi,Chatterjee:2017hya}
and an AB defect \cite{Chatterjee:2019zwx}.

\section{The Alice theory}\label{sec:model}
\subsection{The model}
We consider an $G=SU(2)\times U(1)$ gauge theory coupled with 
one charged complex triplet (adjoint) scalar field 
$\Phi = \sum_{\alpha=1}^3 \Phi^\alpha \sigma^\alpha$ 
and one (charged) doublet scalar field $\Psi$.
Since the matter content is
the same with the gauge and Higgs sectors of the triplet Higgs model beyond the SM, 
we label the gauge group as $G= SU(2)_W\times U(1)_Y$.\footnote{
However we consider a different phase and different parameter region. 
} 
The $U(1)_Y$ charge of  the triplet is $Y = 1$, 
while the doublet fields $q$ ($q=\half$ is the choice of the triplet Higgs model).
In the previous paper \cite{Chatterjee:2019zwx}, 
we considered the cases $q = \half$ and $q=0$,
but here we mostly consider the former.

The Lagrangian is given by
\begin{eqnarray}
\mathcal{L}
&=& -\frac{1}{4} \Tr F_{\mu\nu}F^{\mu\nu} - \frac{1}{4}f_{\mu\nu}f^{\mu\nu} + \Tr | D_\mu\Phi|^2 +  \left|\D_\mu\Psi\right|^2 - V(\Phi,\Psi)
\end{eqnarray}
where 
$
\D_\mu\Phi^\a  = \p_\mu\Phi^\a - ie_{} a_\mu \Phi + g \epsilon^{\alpha\beta\gamma}A_\mu^\beta\Phi^\gamma,\,
 \D_\mu\Psi = \p_\mu \Psi  - i q a_\mu \Psi - i A_\mu \Psi,
 F_{\mu\nu } = \p_\mu A_\nu  - \p_\nu A_\mu- i g [A_\mu, A_\nu],\, f_{\mu\nu } = \p_\mu a_\nu  - \p_\nu a_\mu
$.
Here $e_{}$ and $g$ are the coupling constants of the $U(1)_Y$ and $SU(2)_W$ gauge fields, respectively.
 The potential term is given by
\begin{eqnarray}
\label{potential}
 V(\Phi, \Psi) &=& V_\Phi (\Phi) + V_\Psi(\Psi) + V_{\rm int}(\Phi, \Psi),
\end{eqnarray} 
with 
\begin{eqnarray}
 V_\Phi(\Phi) &=& \frac{\lambda_g}{4} \Tr[\Phi,\Phi^\dagger]^2 +  \frac{\lambda_e}{2}\left(\Tr \Phi\Phi^\dagger - 2 \xi^2\right)^2 , \label{eq:VPhi} \\
 V_\Psi(\Psi) &=&  M^2 \Psi^\dagger\Psi + \lambda_\psi \l(\Psi^\dagger\Psi\r)^2 ,
 \label{eq:VPsi}\\
 V_{\rm int} (\Phi, \Psi) &=& \mu \l( \Psi_c^\dagger \Phi^* \Psi +   \Psi^\dagger \Phi \Psi_c\r)  + 
 \lambda_1 \Psi^\dagger\Psi \Tr \l(\Phi^\dagger\Phi\r)+ \lambda_2 \Psi^\dagger\Phi^\dagger\Phi\Psi, \label{eq:int1} 
\end{eqnarray}
where
 $\lambda_g$ and $\lambda_e$ are two couplings of the triplet, 
  $\xi$ is a parameter giving a VEV of the triplet,  $M$ is the bare mass of the doublet,  $\lambda_\psi$ is the quartic coupling of the doublet field, and
$\mu, \lambda_1, \lambda_2$ are couplings between the doublet and triplet scalar fields.
The charge conjugation of the doublet field
is defined as $\Psi_c = i\sigma^2 \Psi^*$.

\begin{figure}
    \centering
    \includegraphics[width=0.5\columnwidth]{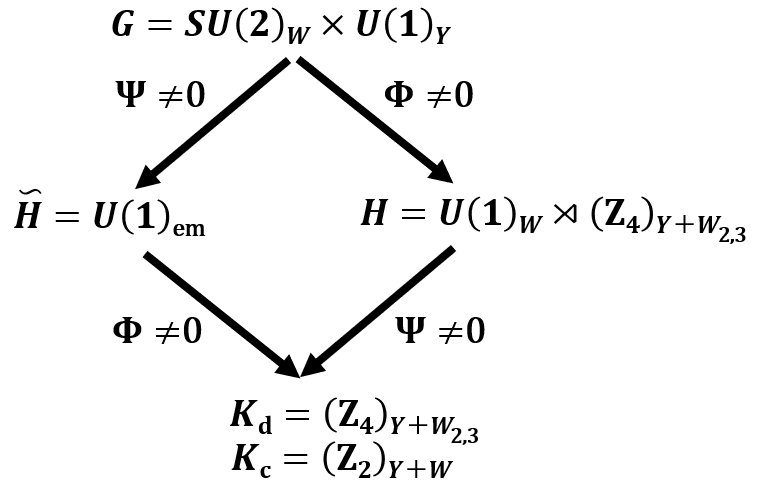}
    \caption{Spontaneous symmetry breaking patterns.}
    \label{fig:SSB}
\end{figure}

Before discussing details in the next section, 
here let us summarize the symmetry breaking patterns 
in Fig.~\ref{fig:SSB} and 
vacuum manifolds and corresponding lower dimensional homotopy groups
in Table~\ref{tab:homo}.

\begin{table}[t]
  \renewcommand{\arraystretch}{0.5}
  \centering
  \begin{tabular}{l|cccc}\hline
    Vacuum manifolds & $\pi_0$ & $\pi_1$ & $\pi_2$ & $\pi_3$
    \\ \hline
    $\dfrac{G}{H} = \frac{U(1)_Y \times SU(2)_W}{(\mathbb{Z}_4)_{Y+W_{2,3}} \ltimes  U(1)_{W_1}  }
\simeq
 \dfrac{S_Y^1 \times S_W^2}{(\mathbb{Z}_2)_{Y+W_{2,3}}}$
     &  & \begin{tabular}{c} $\mathbb{Z}$ \\ Alice
      string \end{tabular} & \begin{tabular}{c} $\mathbb{Z}$ 
      \\ monopole \end{tabular} & $\mathbb{Z}$ \\ \hline
    $\dfrac{H}{K_{\rm d}} \simeq S^1$ &  &     
    \begin{tabular}{c} ~\\$\mathbb{Z}$ \\
   Another string
    \end{tabular}   &  &\\ \hline
    $\dfrac{H}{K_{\rm c}} \simeq  (\mathbb{Z}_2)_{Y+W_{2,3}} \ltimes  U(1)_{W_1}$ 
    & 
    \begin{tabular}{c} 
    $\mathbb{Z}_2$\\ 
    wall 
    \end{tabular}
    & 
    \begin{tabular}{c} ~\\ $\mathbb{Z}$ \\
   Another string
    \end{tabular}  
      &  & 
    \\ \hline\hline
$\dfrac{G}{\tilde H} =  \dfrac{U(1)_Y \times SU(2)_W}{U(1)_{Y+W_3}} 
 \simeq S^3$  
 & & & & 
 ${\mathbb Z}$
 \\ \hline
    $\dfrac{\tilde  H}{K_{\rm d}} = \dfrac{U(1)_{Y+W_3}}{({\mathbb Z}_4)_{Y+W_3}}$ & & \begin{tabular}{c}
      $\mathbb{Z}$\\ $1/4$ Abelian string \end{tabular} & 
     &
     \\ \hline
    $\dfrac{\tilde  H}{K_{\rm c}} = \dfrac{U(1)_{Y+W_3}}{({\mathbb Z}_2)_{Y+W}}$ & & \begin{tabular}{c} $\mathbb{Z}$ \\ \vspace{2pt}
     $1/2$ Abelian string \\  \end{tabular} & &
     \\ \hline\hline
    $\dfrac{G}{K_{\rm d}} 
  =  \dfrac{U(1)_Y\times SU(2)_W}{ (\mathbb{Z}_4)_{Y+ W_{2,3} }} 
  \simeq \dfrac{U(2)_{Y+W}}{({\mathbb Z}_2)_{Y+W_{2,3}}}$
     &  & \begin{tabular}{c} $\mathbb{Z}$\\ $1/4$ Abelian string   \end{tabular} 
     & 
     & ${\mathbb Z}$ \\ \hline
     $\dfrac{G}{K_{\rm c}} 
     =  \dfrac{U(1)_Y \times SU(2)_W}{({\mathbb Z}_2)_{Y+W}} 
 \simeq U(2)_{Y+W}$
     &  & \begin{tabular}{c} $\mathbb{Z}$ \\ $1/2$ Abelian string  \end{tabular} & &${\mathbb Z}$ \\ \hline
  \end{tabular}
  \caption{Vacuum manifolds and corresponding lower dimensional homotopy groups.}
  \label{tab:homo}
\end{table}

\subsection{Symmetry breaking by the triplet field}
First, let us turn off the doublet field 
and consider the triplet field. 
We  choose the vacuum expectation value of the triplet field $\Phi$ without loss of generality as 
\begin{eqnarray}\label{phivac}
\langle\Phi\rangle =  \xi \sigma^1. 
\label{eq:gauge1}
\end{eqnarray}
This triplet VEV breaks the gauge symmetry group spontaneously as\footnote{
When we do not have a doublet field, the actual gauge group is 
$SU(2)/{\mathbb Z}_2 = SO(3)$ since the center 
${\mathbb Z}_2$ does not act on the triplet. 
Even in this case, it is useful to consider the universal covering 
$SU(2)$ for the symmetry breaking.\label{foot:universal-covering}
} 
\begin{eqnarray}
\label{SSB1}
 G =  U(1)_Y \times SU(2)_W  \longrightarrow  
 H = (\mathbb{Z}_4)_{Y+W_{2,3}} \ltimes  U(1)_{W_1} . 
\end{eqnarray}
Here, $U(1)_{W_1}$ is generated by $\sigma^1$, 
and $\ltimes$ denotes a semi-direct product 
implying that 
the $U(1)_{W_1} $ acts on the $(\mathbb{Z}_4)_{Y+W_{2,3}}$ symmetry.  
The $(\mathbb{Z}_4)_{Y+W_{2,3}}$ symmetry 
is defined by a simultaneous action of 
a $\pi$ rotation of the $U(1)_Y$ group  
and an $SU(2)_W$ element generated by
a linear combination of $\sigma^2$ and $\sigma^3$,  
\begin{eqnarray}
 e^{i \frac{\pi}{2} {\bf n} \cdot \vec{\sigma}} 
 = \left(\begin{array}{cc}
  i \cos \beta & \sin \beta \\
 - \sin \beta  & -   i \cos \beta
 \end{array}\right), 
 \quad
 {\bf n} = (0, \sin \beta, \cos \beta),  \label{eq:Z4-in-SU(2)}
\end{eqnarray}
{\it i.~e.}, the minimum element of  $(\mathbb{Z}_4)_{Y+W_{2,3}}$ is given by
\begin{eqnarray}
(\mathbb{Z}_4)_{Y+W_{2,3}}: \quad
 (e^{i \pi},  e^{i \frac{\pi}{2} {\bf n} \cdot \vec{\sigma}}) \in U(1)_Y \times SU(2)_W: \quad
 e^{i \pi}  e^{i \frac{\pi}{2} {\bf n} \cdot \vec{\sigma}} \sigma^1  
 e^{-i \frac{\pi}{2} {\bf n} \cdot \vec{\sigma}} = \sigma^1.
 \label{eq:Z2}
\end{eqnarray}
Because of $(e^{i\pi})^2 =1$ and 
 $(e^{i \frac{\pi}{2} {\bf n} \cdot \vec{\sigma}})^2=-{\bf 1}_2$,
 the double action in Eq.~(\ref{eq:Z2}), 
\begin{eqnarray}
  ({\mathbb Z}_2)_{Y+W} \mbox{ on } \Phi: (1, -{\bf 1}_2) \in U(1)_Y \times SU(2)_W, 
  \label{eq:center-on-Phi}
\end{eqnarray} 
is an element of the center of $SU(2)_W$.\footnote{One has to be careful about the fact that 
 this action behaves as a center only on the triplet field. 
 If there is another field with a non-integer $U(1)_Y$ charge, a non-trivial action of 
 the $U(1)_Y$ action remains on such a field. Our doublet field is such a field, 
 see Eq.~(\ref{eq:center-on-Psi}), below.
 Thus, we write it as $({\mathbb Z}_2)_{Y+W}$ with ``$Y+$.''
 \label{foot:center}
 } 
 We thus have a relation 
\begin{eqnarray}
 && ({\mathbb Z}_4)_{Y+W_{2,3}}  = ({\mathbb Z}_2)_{Y+ W_{2,3}} \times ({\mathbb Z}_2)_{Y+W} .\label{eq:Z4decomp}
\end{eqnarray}
By noting $SU(2)_W/({\mathbb Z}_2)_{Y+W} = SO(3)_W$ 
with the footnote \ref{foot:center},
the vacuum manifold is found to be
\begin{eqnarray}
  \frac{G}{H} &=& \frac{U(1)_Y \times SU(2)_W}{(\mathbb{Z}_4)_{Y+W_{2,3}} \ltimes  U(1)_{W_1}  }
\simeq 
 \frac{U(1)_Y \times [SU(2)_W/({\mathbb Z}_2)_{Y+W}]}{(\mathbb{Z}_2)_{Y+W_{2,3}} \ltimes  U(1)_{W_1}  } 
\simeq
 \frac{S_Y^1 \times S_W^2}{(\mathbb{Z}_2)_{Y+W_{2,3}}}.
 \label{eq:OPS}
\end{eqnarray}
In the rightmost expression, 
the $({\mathbb Z}_2)_{Y+W_{2,3}}$ symmetry is generated by the first factor of the right hand side of Eq.~(\ref{eq:Z4decomp}).

The lower dimensional homotopy groups for this manifold 
can be calculated as
\begin{eqnarray}
 \pi_0  \left( \frac{G}{H}\right)
\simeq \{0\}, 
\quad
 \pi_1  \left( \frac{G}{H}\right)
\simeq \mathbb{Z},
\quad
\pi_2  \left( \frac{G}{H}\right)
\simeq \mathbb{Z},
\quad
\pi_3  \left( \frac{G}{H}\right)
\simeq \mathbb{Z},\label{eq:homotopy-G/H}
\end{eqnarray}
indicating the absence of domain walls, 
the existence of stable strings (vortices), 
and monopoles.\footnote{
From the nontrivial third homotopy group $\pi_3$, Hopfions are possible 
at least for global analogues \cite{Kawaguchi:2008xi}, 
but it is unclear whether gauged Hopfions stably exist for the local case.
}

\subsection{Symmetry breaking by the doublet field}
Next, let us turn on the doublet field 
supposing that the triplet field develops VEVs at high energy 
and the doublet develops VEVs at low energy. 
The $U(1)_{W_1}$ group in $H$ in Eq.~(\ref{SSB1}) must be spontaneously broken by the doublet VEVs, but the breaking pattern of 
the $(\mathbb{Z}_4)_{Y+W_{2,3}}$ symmetry depends on the VEVs.
The $(\mathbb{Z}_4)_{Y+W_{2,3}}$ symmetry in Eq.~(\ref{eq:Z2}) acts on the doublet field 
$\Psi = (a,b)^T$ ($|a|^2+|b|^2=v^2$) with the $U(1)_Y$ charge $q$ as
\begin{eqnarray}
 \Psi' 
=  e^{i \pi q} \left(\begin{array}{cc}
  i \cos \beta & \sin \beta \\
 - \sin \beta  & -   i  \cos \beta
 \end{array}\right) 
 \left(\begin{array}{c} 
 a \\ b
 \end{array}\right)
 =  e^{i \pi q}
  \left(\begin{array}{c} 
  i a  \cos \beta + b  \sin \beta \\
 -i b  \cos \beta  - a \sin \beta
 \end{array}\right). \label{eq:Z2-cond}
\end{eqnarray}
From now on, we consider the case of $q=1/2$.

We call the case in which 
the $(\mathbb{Z}_4)_{Y+W_{2,3}}$ symmetry in Eq.~(\ref{eq:Z2}) is unbroken 
as the deconfined phase, 
and otherwise the confined phase. 
Two typical solutions of the deconfined phase are
\begin{eqnarray}
\left\{
\begin{array}{ccc}
  \Psi = 
 \left(
\begin{array}{c}
  0  \\
  v
\end{array}
\right) ,
 &  \cos \beta =+1 ,  & K = (\mathbb{Z}_4)_{Y+ W_{3}}\cr 
  \Psi = 
 \left(
\begin{array}{c}
  v  \\
  0
\end{array}
\right) ,  &  \cos \beta =-1, & K = (\mathbb{Z}_4)_{Y- W_{3}},
\end{array}\right.
 \label{eq:A-B0}
\end{eqnarray}
where $(\mathbb{Z}_4)_{Y+W_{2,3}}$ in 
Eqs.~(\ref{eq:Z4-in-SU(2)}) and (\ref{eq:Z2})
is generated by a $\pi$ rotation of $U(1)$ and $+ \sigma^3$ or $-\sigma^3$ in the $SU(2)$ group, respectively,  
and so we have denoted it by $(\mathbb{Z}_4)_{Y\pm W_{3}}$.  
More general solution of the deconfined phase is found to be
\begin{eqnarray}
&& \Psi = U  \left(
\begin{array}{c}
  0  \\
  v
\end{array}
\right) 
=  \left(
\begin{array}{c}
  i v \sin \frac{\beta}{2}  \\
   v \cos \frac{\beta}{2}
\end{array}
\right) , \quad 
K = ({\mathbb Z_4})_{Y+W_{2,3}},
 \label{eq:A-B}
\end{eqnarray}
with $({\mathbb Z}_4)_{Y+W_{2,3}}$ 
in Eq.~(\ref{eq:Z2}) 
and $U$ defined by
\begin{eqnarray}
&& U = e^{\frac{i}{2} \beta \sigma^1} 
      = \cos \frac{\beta}{2} {\bf 1}_2 +  i \sigma^1 \sin \frac{\beta}{2} 
      =  \left(
\begin{array}{ccc}
    \cos \frac{\beta}{2} & i \sin \frac{\beta}{2}      \\
  i \sin \frac{\beta}{2}  &  \cos \frac{\beta}{2}
\end{array}
\right).\label{eq:U-sigma1}
\end{eqnarray}
Here, $\beta=0$ and $\beta=\pi$ correspond to the two cases in Eq.~(\ref{eq:A-B0}) (up to the constant factor $i$), respectively. 
The parameter $\beta$ corresponds to the $U(1)_{W_1}$ (would-be) Nambu-Goldstone mode 
associated with the $U(1)_{W_1}$ symmetry 
spontaneously broken by the doublet.

From the $(\mathbb{Z}_4)_{Y+W_{2,3}}$ 
action on $\Psi$ in Eq.~(\ref{eq:Z2-cond}) 
with $q=1/2$, the double action, 
\begin{eqnarray} 
  (\mathbb{Z}_2)_{Y+W}  \mbox{ on } \Psi : (-1,-{\bf 1}_2) \in U(1)_Y \times SU(2)_W,
  \label{eq:center-on-Psi}
\end{eqnarray}
always leaves arbtrary $\Psi$ invariant, 
thanks to the particular choice of $q=1/2$.
As noted in footnote \ref{foot:center}, this $ (\mathbb{Z}_2)_{Y+W}$ action behaves as a center of $SU(2)_W$ when acting only on the triplet field $\Phi$ as 
in Eq.~(\ref{eq:center-on-Phi}) but it is not so on the doublet field $\Psi$ 
because of the $-1$ element of $U(1)_Y$.
This fact becomes important in the vacuum manifolds.
Thus,  in the confined phase, the $(\mathbb{Z}_4)_{Y+W_{2,3}}$ 
is spontaneously broken 
to a $(\mathbb{Z}_2)_{Y+W}$ subgroup.\footnote{
For different choice of $q$, this $(\mathbb{Z}_2)_{Y+W}$ 
is broken in general.
}

One can easily verify 
that the configurations in Eq.~(\ref{eq:A-B}) satisfy the gauge invariant (covariant) condition  
\begin{eqnarray} 
\Psi^\dagger\Phi\Psi = 0.
\end{eqnarray}
Thus, the phases and unbroken symmetries can be summarized as
\begin{eqnarray}
  \begin{array}{ccc}
   \mbox{deconfined phase: }&    \Psi^\dagger\Phi\Psi = 0, & 
 \;\;  K_{\rm d} = (\mathbb{Z}_4)_{Y+ W_{2,3}}, \cr
   \mbox{confined phase: } &  |\Psi^\dagger\Phi\Psi| \neq 0, & 
   K_{\rm c} = (\mathbb{Z}_2)_{Y+W},
  \end{array}
  \label{eq:A-B2}
\end{eqnarray}
for the deconfined and confined phases, respectively.

The vacuum manifolds for the second symmetry breaking are
\begin{eqnarray}
 \mbox{deconfined phase: } &&
 \frac{H}{K_{\rm d}} = S^1 ,
 \nonumber\\
 \mbox{confined phase: } &&
 \frac{H}{K_{\rm c}} =  (\mathbb{Z}_2)_{Y+W_{2,3}} \ltimes  U(1)_{W_1} .%
\end{eqnarray}
For the both cases, the $S^1\simeq U(1)_{W_1}$ part of the vacuum manifolds is parameterized by $\beta$ in Eq.~(\ref{eq:A-B}).
The associated lower dimensional 
homotopy groups for the second symmetry breaking are 
\begin{eqnarray}
 \mbox{deconfined
 : }
&& \pi_0  \left( \frac{H}{K_{\rm d}}\right)
\simeq \{0\} , 
\quad
 \pi_1  \left( \frac{H}{K_{\rm d}}\right)
\simeq \mathbb{Z},
\quad
 \pi_2  \left( \frac{H}{K_{\rm d}}\right)
\simeq  \{0\},
\quad
 \pi_3  \left( \frac{H}{K_{\rm d}}\right)
\simeq  \{0\} ,  \nonumber \\
 \mbox{confined 
 : }
&& \pi_0  \left( \frac{H}{K_{\rm c}}\right)
\simeq \mathbb{Z}_2,
\quad
 \pi_1  \left( \frac{H}{K_{\rm c}}\right)
\simeq \mathbb{Z},
\quad
 \pi_2  \left( \frac{H}{K_{\rm c}}\right)
\simeq  \{0\},
\quad
 \pi_3  \left( \frac{H}{K_{\rm c}}\right)
\simeq  \{0\}. \nonumber\\ \label{eq:homotopy-doublet}
\end{eqnarray}
These show that both the phases allow another type of vortices, 
and that 
the confined phase allows ${\mathbb Z}_2$ domain walls while 
the deconfined phase does not.

Finally,
the vacuum manifolds of the overall breaking by the both triplet and doublet fields 
are 
\begin{eqnarray}
 \mbox{deconfined phase: } && \frac{G}{K_{\rm d}} 
  =  \frac{U(1)_Y\times SU(2)_W}{ (\mathbb{Z}_4)_{Y+ W_{2,3} }}
 \simeq 
 \dfrac{\left( \frac{U(1)_Y \times SU(2)_W}{({\mathbb Z}_2)_{Y+W}} \right)}{(\mathbb{Z}_2)_{Y+ W_{2,3}}}
  \simeq \frac{U(2)_{Y+W}}{({\mathbb Z}_2)_{Y+W_{2,3}}} 
 , \nonumber\\
 \mbox{confined phase: } && \frac{G}{K_{\rm c}} =  \frac{U(1)_Y \times SU(2)_W}{({\mathbb Z}_2)_{Y+W}} 
 \simeq U(2)_{Y+W} , 
 \label{eq:G/K}
\end{eqnarray}
for the deconfined and confined phases, respectively.\footnote{
It is interesting to note that, 
eventually, the vacuum manifold of the overall breaking in the confined phase 
coincides with that of the $U(2)$ Higgs model admitting non-Abelian vortices 
\cite{Hanany:2003hp,Auzzi:2003fs,
Eto:2004rz,
Eto:2005yh,Eto:2006cx,
Tong:2005un,Eto:2006pg,Shifman:2007ce,Shifman:2009zz}.
} 
The lower dimensional homotopy groups are 
\begin{eqnarray}
 \mbox{deconfined: }
&& \pi_0  \left( \frac{G}{K_{\rm d}}\right)
\simeq \{0\},
\quad
 \pi_1  \left( \frac{G}{K_{\rm d}}\right)
\simeq \mathbb{Z},
\quad
 \pi_2  \left( \frac{G}{K_{\rm d}}\right)
\simeq  \{0\},
\quad
 \pi_3  \left( \frac{G}{K_{\rm d}}\right)
\simeq   {\mathbb Z} , \nonumber \\
 \mbox{confined: }
&& \pi_0  \left( \frac{G}{K_{\rm c}}\right)
\simeq \{0\},
\quad
 \pi_1  \left( \frac{G}{K_{\rm c}}\right)
\simeq \mathbb{Z}  ,
\quad
 \pi_2  \left( \frac{G}{K_{\rm c}}\right)
\simeq  \{0\},
\quad
 \pi_3  \left( \frac{G}{K_{\rm c}}\right)
\simeq  {\mathbb Z}. \nonumber\\ \label{eq:homotopy-overall}
\end{eqnarray}
The both phases have nontrivial first homotopy groups. 
However,  the minimal elements are different
as it will become important later.

\bigskip
For the case of $q=0$, 
the condition $\Psi'=\Psi=(a,b)^T$
with Eq.~(\ref{eq:Z2-cond}) 
cannot be satisfied,
implying that the ${\mathbb Z}_4$ symmetry is broken 
completely
by the VEV of the doublet 
(where no ${\mathbb Z}_2$ remians).

\subsection{The opposite hierarchy}\label{sec:opposite}

Although we mainly consider the case that 
the triplet VEVs are much greater than the doublet VEVs,
let us briefly mention the opposite case 
(with the doublet $U(1)_Y$ charge $q=1/2$),  
in which the doublet VEVs are much greater than 
the triplet VEVs. 
As the case of the SM,
when the doublet field develops VEVs, we can take
\begin{eqnarray}
 \Psi = (v,0)^T \label{eq:Psi-vev}
\end{eqnarray}
without loss of generality.
The symmetry is broken as\footnote{
In this case, the actual gauge group is 
$[U(1)_Y \times SU(2)_W ] /({\mathbb Z}_2)_{Y+W}$, but it is useful to consider the universal covering 
$U(1)_Y \times SU(2)_W$. 
The vacuum manifold 
in Eq.~(\ref{eq:vacuum-mfd-doublet}) below 
is the same.
See also the footnote \ref{foot:universal-covering}.
}
\begin{eqnarray}
\label{SSB2}
 G =  U(1)_Y \times SU(2)_W  \longrightarrow 
 \tilde  H = U(1)_{Y+W_3} [= U(1)_{\rm em} ] 
\end{eqnarray}
where $U(1)_{Y+W_3}$ is defined by
\begin{eqnarray} 
U(1)_{Y+W_3}: \; \Psi \to e^{- i \alpha} e^{i \alpha \sigma^3} \Psi,\label{eq:U(1)em}
\end{eqnarray}
which is the electromagnetic $U(1)_{\rm em}$ 
in the context of the SM.
The vacuum manifold is
\begin{eqnarray}
 \frac{G}{\tilde H} =  \frac{U(1)_Y \times SU(2)_W}{U(1)_{Y+W_3}} 
 \simeq SU(2) \simeq S^3,\label{eq:vacuum-mfd-doublet}
\end{eqnarray}
with lower dimensional homotopy groups
\begin{eqnarray}
 \pi_0  \left( \frac{G}{\tilde H}\right)
\simeq \{0\}, 
\quad
 \pi_1  \left( \frac{G}{\tilde H}\right)
\simeq \{0\}, 
\quad
\pi_2  \left( \frac{G}{\tilde H}\right)
\simeq \{0\}, 
\quad
\pi_3  \left( \frac{G}{\tilde H}\right)
\simeq \mathbb{Z},\label{eq:homotopy-G/tilde-H}
\end{eqnarray}
which are topologically almost  trivial as is well known.

Then, when the triplet develops VEVs at low energy, 
there are two phases as summarized in Eq.~(\ref{eq:A-B2}):
\begin{eqnarray}
 \mbox{deconfined phase: } && \Psi^\dagger \Phi \Psi = 0, \quad
 \Phi = \xi (\cos \gamma \sigma^1 + \sin \gamma \sigma^2)
 =  \xi \left(
\begin{array}{cc}
                    0 &  e^{ i \gamma}     \\
    e^{- i \gamma}  & 0  
\end{array}
\right) ,\nonumber\\
 \mbox{confined phase: } && \Psi^\dagger \Phi \Psi \neq 0, \quad \Phi = \xi \sigma^3,   \label{eq:Phi-VEV}
\end{eqnarray}
for Eq.~(\ref{eq:Psi-vev}).
The unbroken symmetry 
$\tilde  H = U(1)_{Y+W_3}$ is further broken to 
either ${\mathbb Z}_4$ or ${\mathbb Z}_2$ 
(unlike the triplet Higgs model beyond the SM 
in which $\tilde H$ should be unbroken).
In the deconfined phase, $\Phi$ is invariant under the ${\mathbb Z}_4$ symmetry whose minimum element is given by
\begin{eqnarray}
(\mathbb{Z}_4)_{Y+W_{3}}: \quad
 (e^{i \pi},  e^{i \frac{\pi}{2} \sigma^3 }) \in U(1)_Y \times SU(2)_W: \quad
 e^{i \pi}  e^{i \frac{\pi}{2} \sigma^3} \Phi  
 e^{-i \frac{\pi}{2} \sigma^3} = \Phi.
 \label{eq:Z2-opposite}
\end{eqnarray}
In the confined phase, $\Phi = \xi \sigma^3$ is invariant under 
the double action of Eq.~(\ref{eq:Z2-opposite}) 
generating $({\mathbb Z}_2)_{Y+W}$ in Eq.~(\ref{eq:center-on-Psi}).
Thus, we have unbroken symmetries 
\begin{eqnarray}
  \mbox{deconfined phase: } && K_{\rm d}= ({\mathbb Z}_4)_{Y+W_3}, \nonumber\\ 
  \mbox{confined phase: } && K_{\rm c} = ({\mathbb Z}_2)_{Y+W}  ,
\end{eqnarray}
consistent with the previous discussion as it should be so. 
One can easily check that these discrete groups are subgroups of 
the ${U(1)_{Y+W_3}}$ group in Eq.~(\ref{eq:U(1)em}).
Thus, the vacuum manifolds for the second symmetry breaking are found to be
\begin{eqnarray}
  \mbox{deconfined phase: } && \frac{\tilde  H}{K_{\rm d}} = \frac{U(1)_{Y+W_3}}{({\mathbb Z}_4)_{Y+W_3}}, \nonumber\\ 
  \mbox{confined phase: } &&
 \frac{\tilde  H}{K_{\rm c}} = \frac{U(1)_{Y+W_3}}{({\mathbb Z}_2)_{Y+W}}, 
 \label{eq:opposite-vacuum-mfd}
\end{eqnarray}
and the lower dimensional homotopy groups are  
\begin{eqnarray}
 \mbox{deconfined: }
&& \pi_0  \left( \frac{\tilde H}{K_{\rm d}}\right)
\simeq \{0\} ,
\quad
 \pi_1  \left( \frac{\tilde H}{K_{\rm d}}\right)
\simeq \mathbb{Z},
\quad
 \pi_2  \left( \frac{\tilde H}{K_{\rm d}}\right)
\simeq  \{0\},
\quad
 \pi_3  \left( \frac{\tilde H}{K_{\rm d}}\right)
\simeq  \{0\} ,  \nonumber \\
 \mbox{confined: }
&& \pi_0  \left( \frac{\tilde H}{K_{\rm c}}\right)
\simeq   \{0\},
\quad
 \pi_1  \left( \frac{\tilde H}{K_{\rm c}}\right)
\simeq \mathbb{Z},
\quad
 \pi_2  \left( \frac{\tilde H}{K_{\rm c}}\right)
\simeq  \{0\},
\quad
 \pi_3  \left( \frac{\tilde H}{K_{\rm c}}\right)
\simeq  \{0\}. \nonumber\\ \label{eq:homotopy-opposite}
\end{eqnarray}
There are only nontrivial first homotopy groups 
admitting strings, 
which look the same for the deconfined and confined phases. However, 
the vacuum manifolds in Eq.~(\ref{eq:opposite-vacuum-mfd}) 
show that the deconfined and confined phases 
 allow 1/4 quantized and 1/2 quantized strings, respectively.

When both the triplet and doublet fields develop VEVs, 
the unbroken symmetries and physics 
are the same between the two cases 
that the triplet (doublet) develops VEVs at high energy 
and the doublet (triplet) develops VEVs at low energy.
In the above discussion, these two cases look different at first glance;
for instance, in the former,
the doublet field $\Psi$ contains a continuous parameter $\beta$ in Eq.~(\ref{eq:A-B}), 
and in the lattter, 
the triplet field $\Phi$ contains a continuous parameter $\gamma$ in Eq.~(\ref{eq:Phi-VEV}).
This is just because of a difference of the gauge fixings: 
for the former (latter) 
the gauge fixing in Eq.~(\ref{eq:gauge1}) [(\ref{eq:Psi-vev})] 
was convenient, 
and these two gauges can be of course transformed to each other 
by a gauge transformation.
In fact, the classification of the deconfined and confined phase is the identical 
in a gauge invariant description, 
as can be seen in Eq.~(\ref{eq:A-B2}),  
 
\section{Strings and monopoles}\label{sec:string-monopole}
In this section, we discuss string and monopole configurations associated with homotopy groups 
in Eq.~(\ref{eq:homotopy-G/H}) 
in the absence of the doublet field. 

\subsection{Abrikosov-Nielsen-Olesen string and ${\mathbb Z}_2$ string}\label{sec:ANO-Z2}

The ANO string \cite{Abrikosov:1956sx,Nielsen:1973cs} 
is a vortex having a phase winding only in the $U(1)$ gauge group.
In our case, it has the form of  
\begin{eqnarray}
&& \Phi (r, \theta)  
\sim 
 \xi \left(
\begin{array}{ccc}
  0 &  e^{ i \theta}     \\
    e^{ i \theta}  & 0  
\end{array}
\right) 
=   e^{ i \theta} \sigma^1,
\nonumber  \\
&& A_i = 0 ,
\quad a_i \sim - \frac{1}{ e_{}}\frac{\epsilon_{ij} x^j}{r^2}.\label{eq:ANO}
\end{eqnarray}
This has a unit $U(1)$ magnetic flux 
and no $SU(2)$ flux. 
This is BPS when we take the BPS limit $\lambda_e = e$ and $\lambda_g=g$.

\bigskip
Next, we consider
a ${\mathbb Z}_2$ string 
which exists in an $SU(2)$ gauge theory with a complex adjoint field
\cite{Abrikosov:1956sx}.\footnote{
In the original case \cite{Abrikosov:1956sx}, 
$U(1)$ is a global symmetry.}
It has the form of
\begin{eqnarray}
\label{Aliceconfiguration-Z2}
&& \Phi(r, \theta)  \sim 
\xi \left(
\begin{array}{ccc}
 0 &  e^{i\theta}     \\
  e^{-i\theta} & 0  
\end{array}
\right) 
= \xi e^{i\frac{\theta}{2}\sigma^3} \sigma^1 e^{-i\frac{\theta}{2}\sigma^3}, \nonumber 
\\
&& 
A_i \sim - \frac{1}{2 g}\frac{\epsilon_{ij} x^j}{r^2} \sigma^3, \quad 
a_i = 0.
  \label{eq:Alice-asymptotic-Z2}
\end{eqnarray}
The $SU(2)$ group element $e^{i\frac{\theta}{2}\sigma^3}$ 
varies from $1$ to $-1$ when $\theta$ changes from $\theta = 0$ to $\theta =2\pi$.
Thus, the ${\mathbb Z}_2$ string connects the center elements $\pm 1$ of $SU(2)$
and carries a half amount of the $SU(2)$ magnetic flux 
compared with one corresponding to a closed loop in $SU(2)$.\footnote{
Since the center does not act on the triplet in 
the adjoint action, 
the actual gauge group is 
$SO(3) = SU(2) /{\mathbb Z}_2$ 
unless we introduce a doublet field. 
In this case, 
Eq.~(\ref{eq:Alice-asymptotic-Z2}) gives a closed loop in $SO(3)$. 
While 
it carries a half amount of flux in terms of $SU(2)$, 
it carries a {\it unit} flux in terms of $SO(3)$ because it is given by a closed loop 
in $SO(3)$.
In this paper, however, we also consider the doublet, and so our gauge group is 
$SU(2)$ rather than $SO(3)$. 
}
This is non-BPS. However, it is stable.

\subsection{Alice strings}\label{sec:Alice-AB}

Let us introduce an Alice string discussed in 
Refs.~\cite{Chatterjee:2017jsi,Chatterjee:2017hya,Chatterjee:2019zwx}.
  The configuration of  
  an Alice string  can be expressed at a large distance as
\begin{eqnarray}
\label{Aliceconfiguration}
&& \Phi(r, \theta)  
\sim 
 \xi \left(
\begin{array}{ccc}
  0 &  e^{ i \theta}     \\
  1 & 0  
\end{array}
\right) = 
\xi e^{i\frac{\theta}{2}} \left(
\begin{array}{ccc}
                            0 &  e^{ + i\frac{\theta}{2}}     \\
  e^{-i\frac{\theta}{2}} & 0  
\end{array}
\right) 
= \xi e^{i\frac{\theta}{2}} e^{+i\frac{\theta}{4}\sigma^3} \sigma^1 e^{-i\frac{\theta}{4}\sigma^3}, \nonumber 
\\
&& A_i \sim - \frac{1}{4 g}\frac{\epsilon_{ij} x^j}{r^2} \sigma^3, \quad 
a_i \sim - \frac{1}{2 e_{}}\frac{\epsilon_{ij} x^j}{r^2},
  \label{eq:Alice-asymptotic}
\end{eqnarray}
or
\begin{eqnarray}
\label{Aliceconfiguration2}
&& \Phi(r, \theta)  
\sim  \xi \left(
\begin{array}{ccc}
                 0 &  1   \\
  e^{ i \theta}  & 0  
\end{array}
\right) 
= \xi e^{i\frac{\theta}{2}} \left(
\begin{array}{ccc}
                            0 &  e^{-i\frac{\theta}{2}}     \\
  e^{+i\frac{\theta}{2}} & 0  
\end{array}
\right) 
= \xi e^{i\frac{\theta}{2}} e^{-i\frac{\theta}{4}\sigma^3} \sigma^1 e^{+i\frac{\theta}{4}\sigma^3}, \nonumber 
\\
&& A_i \sim + \frac{1}{4 g}\frac{\epsilon_{ij} x^j}{r^2} \sigma^3, \quad 
a_i \sim - \frac{1}{2 e_{}}\frac{\epsilon_{ij} x^j}{r^2}
  \label{eq:Alice-asymptotic2}
\end{eqnarray}
with an angular coordinate $\theta$. 
It carries a half amount of the $U(1)$ magnetic flux and $1/4$ amount of 
the $SU(2)$ color magnetic flux, 
taking a value on $\pm \sigma^3$ in this case. 
It is important to note that the 
$1/4$ quantization of the $SU(2)$ color magnetic flux 
of the Alice string 
is a consequence of $(\mathbb{Z}_4)_{Y+W_{2,3}}$ in Eq.~(\ref{eq:OPS}).
The Alice string also can be regarded as a half ANO string as well as 
a half ${\mathbb Z}_2$ string.

A full configuration including finite $r$ can be
obtained by complementing profile functions 
in the asymptotic form in Eq.~(\ref{eq:Alice-asymptotic}).
See Appendix \ref{sec:app}.

These two configurations are merely typical configurations 
belonging to a continuous family of 
more general configurations, given by 
\begin{eqnarray}
\label{Aliceconfiguration3}
&& \Phi(r, \theta)  \sim 
 \xi U \left(
\begin{array}{ccc}
  0 &  e^{ i \theta}     \\
  1 & 0  
\end{array}
\right) U^\dagger 
= 
\xi e^{i\frac{\theta}{2}} e^{+i\frac{\theta}{4}\hat \sigma} \sigma^1 e^{-i\frac{\theta}{4}\hat\sigma}
, \nonumber 
\\
&& A_i \sim - \frac{1}{4 g}\frac{\epsilon_{ij} x^j}{r^2} \hat \sigma, \quad 
a_i \sim - \frac{1}{2 e_{}}\frac{\epsilon_{ij} x^j}{r^2}, 
    \label{eq:general-Alice}
\end{eqnarray}
where  $U$ is defined in Eq.~(\ref{eq:U-sigma1})
and $\hat \sigma$ is defined by
\begin{eqnarray}
 \hat\sigma 
   = U \sigma^3 U^\dagger
  = \sin \beta \sigma^2 + \cos \beta \sigma^3 .
      \label{eq:hat-sigma}
\end{eqnarray}
Here, $\beta$ is a $U(1)$ modulus characterizing an $SU(2)$ color magnetic flux 
between $\sigma^2$ and $\sigma^3$,
and $\beta =0$ ($\beta =\pi$) corresponds to 
the configuration in Eq.~(\ref{eq:Alice-asymptotic})
[Eq.~(\ref{eq:Alice-asymptotic2})].

On the other hand, 
anti-strings carrying
a $-1/2$ $U(1)$ magnetic flux 
and $1/4$ $SU(2)$ magnetic flux
can be constructed as
\begin{eqnarray}
\label{Aliceconfiguration-anti}
&& \Phi(r, \theta)  
\sim 
 \xi \left(
\begin{array}{ccc}
                        0 & 1     \\
    e^{ -i \theta}     & 0  
\end{array}
\right) = 
\xi e^{- i\frac{\theta}{2}} \left(
\begin{array}{ccc}
                            0 &  e^{ + i\frac{\theta}{2}}     \\
  e^{-i\frac{\theta}{2}} & 0  
\end{array}
\right) 
= \xi e^{- i\frac{\theta}{2}} e^{+i\frac{\theta}{4}\sigma^3} \sigma^1 e^{-i\frac{\theta}{4}\sigma^3}, \nonumber 
\\
&& A_i \sim - \frac{1}{4 g}\frac{\epsilon_{ij} x^j}{r^2} \sigma^3, \quad 
a_i \sim + \frac{1}{2 e_{}}\frac{\epsilon_{ij} x^j}{r^2},
  \label{eq:Alice-asymptotic-anti}
\end{eqnarray}
or
\begin{eqnarray}
\label{Aliceconfiguration2-anti}
&& \Phi(r, \theta)  
\sim  \xi \left(
\begin{array}{ccc}
  0 &    e^{- i \theta}     \\
  1 & 0  
\end{array}
\right) 
= \xi e^{-i\frac{\theta}{2}} \left(
\begin{array}{ccc}
                            0 &  e^{-i\frac{\theta}{2}}     \\
  e^{+i\frac{\theta}{2}} & 0  
\end{array}
\right) 
= \xi e^{-i\frac{\theta}{2}} e^{-i\frac{\theta}{4}\sigma^3} \sigma^1 e^{+i\frac{\theta}{4}\sigma^3}, \nonumber 
\\
&& A_i \sim + \frac{1}{4 g}\frac{\epsilon_{ij} x^j}{r^2} \sigma^3, \quad 
a_i \sim + \frac{1}{2 e_{}}\frac{\epsilon_{ij} x^j}{r^2}.
  \label{eq:Alice-asymptotic2-anti}
\end{eqnarray}
Again, more generally, they belong to a continuous family of configuration:
\begin{eqnarray}
\label{Aliceconfiguration3}
&& \Phi(r, \theta)  \sim 
\xi U \left(
\begin{array}{ccc}
                        0 & 1     \\
    e^{ -i \theta}     & 0  
\end{array}
\right) U^\dagger = 
\xi e^{-i\frac{\theta}{2}} e^{+i\frac{\theta}{4}\hat \sigma} \sigma^1 e^{-i\frac{\theta}{4}\hat\sigma}, \nonumber 
\\
&& A_i \sim - \frac{1}{4 g}\frac{\epsilon_{ij} x^j}{r^2} \hat \sigma, \quad 
a_i \sim + \frac{1}{2 e_{}}\frac{\epsilon_{ij} x^j}{r^2},
  \label{eq:Alice-asymptotic3-anti} 
\end{eqnarray}
with 
$U$ defined in Eq.~(\ref{eq:U-sigma1}) and 
$\hat \sigma$ defined in Eq.~(\ref{eq:hat-sigma}).

Let us briefly show the Alice property.
The triplet field approaches 
$\Phi(r \to \infty, \theta=0) = \xi \sigma^1$ asymptotically 
along the $x^1$-axis ($\theta=0$), 
and the $\theta$-dependence is
 is given by 
\begin{align}
\Phi(r \to \infty, \theta) &\sim e^{ie_{} \int {\bf a\cdot dl}} Pe^{i g \int {\bf A\cdot dl}}\, \Phi(r \to \infty, 0)Pe^{-ig \int {\bf A\cdot dl}} \nonumber\\
 & \sim U_0(\theta) U_3(\theta)\, \Phi(r \to \infty, 0) U^{-1}_3(\theta),\;
\end{align}
where the holonomies are defined as
\begin{eqnarray}
\label{holonomy}
U_0(\theta) = e^{ie \int_0^\theta {\bf a\cdot dl}} = e^{i\frac{\theta}{2}}\in U(1), \quad U_3(\theta) = Pe^{i g \int_0^\theta {\bf A\cdot dl}} = e^{i\frac{\theta}{4}\sigma^3} \in SU(2).
\end{eqnarray}
These can be understood by using the condition of topological vortex where  the order parameter $\Phi$ is covariantly constant at large distances ($\D_i\Phi \rightarrow 0$ as $r \rightarrow \infty$).  
When we encircle the string by an angle $\theta$,   the unbroken $U(1)$  generator $Q$ transforms as 
\begin{eqnarray}
Q_\theta = U_3(\theta) Q_0 U_3(\theta)^{-1}, 
\end{eqnarray}
with 
$Q_0 = \sigma^1$. 
We thus have a sign flip of the $U(1)$ generator after one complete encirclement as
 \begin{eqnarray}
Q_{2\pi} = - Q_0
\end{eqnarray}
implying that the unbroken generator is not singly defined 
around the string, 
which is nothing but the Alice property 
and sometimes called a topological obstruction.

For later convenience, 
let us give a doubly-wound Alice string configuration, 
which is 
the most important 
among multiple Alice string configurations.
It has the form of
\begin{eqnarray}
&& \Phi(r, \theta)  \sim 
\xi \left(
\begin{array}{ccc}
 0 &  e^{2i\theta}     \\
 1 & 0  
\end{array}
\right) =
\xi e^{i \theta} \left(
\begin{array}{ccc}
 0 &  e^{+i\theta}     \\
  e^{-i\theta} & 0  
\end{array}
\right) 
= \xi e^{i\theta} e^{i\frac{\theta}{2}\sigma^3} \sigma^1 e^{-i\frac{\theta}{2}\sigma^3}, \nonumber 
\\
&& A_i \sim - \frac{1}{2 g}\frac{\epsilon_{ij} x^j}{r^2} \sigma^3, \quad 
a_i \sim - \frac{1}{ e_{}}\frac{\epsilon_{ij} x^j}{r^2}
  \label{eq:Alice-asymptotic-double}
\end{eqnarray}
or 
\begin{eqnarray}
&& \Phi(r, \theta)  \sim 
\xi \left(
\begin{array}{ccc}
 0 & 1     \\
  e^{2i\theta} & 0  
\end{array}
\right) =
\xi e^{i \theta} \left(
\begin{array}{ccc}
 0 &  e^{-i\theta}     \\
  e^{+i\theta} & 0  
\end{array}
\right) 
= \xi e^{i\theta} e^{i\frac{\theta}{2}\sigma^3} \sigma^1 e^{-i\frac{\theta}{2}\sigma^3}, \nonumber 
\\
&& A_i \sim + \frac{1}{2 g}\frac{\epsilon_{ij} x^j}{r^2} \sigma^3, \quad 
a_i \sim - \frac{1}{ e_{}}\frac{\epsilon_{ij} x^j}{r^2}
  \label{eq:Alice-asymptotic-double2}
\end{eqnarray}
for a specific orientation and 
\begin{eqnarray}
&& \Phi(r, \theta)  \sim 
\xi U \left(
\begin{array}{ccc}
 0 &  e^{2i\theta}     \\
 1 & 0  
\end{array}
\right) U^\dagger
=
\xi e^{i \theta} e^{+i\frac{\theta}{2}\hat \sigma} \sigma^1 e^{-i\frac{\theta}{2}\hat\sigma}, \nonumber 
\\
&& A_i \sim - \frac{1}{2 g}\frac{\epsilon_{ij} x^j}{r^2} \hat \sigma, \quad 
a_i \sim - \frac{1}{ e_{}}\frac{\epsilon_{ij} x^j}{r^2}, 
  \label{eq:general-Alice-double}
\end{eqnarray}
with 
$U$ defined in Eq.~(\ref{eq:U-sigma1}) and 
$\hat \sigma$ defined in Eq.~(\ref{eq:hat-sigma}),
for more general configuration. 
Here, $\beta$ is again a $U(1)$ modulus characterizing an $SU(2)$ color magnetic flux 
between $\sigma^2$ and $\sigma^3$.
Each carries a unit $U(1)$ magnetic flux and a half $SU(2)$ magnetic flux, 
which are just twice of a single Alice string.
This can also be regarded as 
a composite of one ANO vortex and one ${\mathbb Z}_2$ vortex.
This should be BPS, since it is just twice of one Alice string 
(although a ${\mathbb Z}_2$ string is not BPS). 
The unbroken generator $Q$ is singly defined around 
the doubly-wound Alice string,
unlike a single Alice string.


\subsection{Alice monopoles}\label{sec:monopole}
\begin{figure}[!htb]
\centering
\begin{tabular}{cc}
\includegraphics[totalheight=4.5cm]{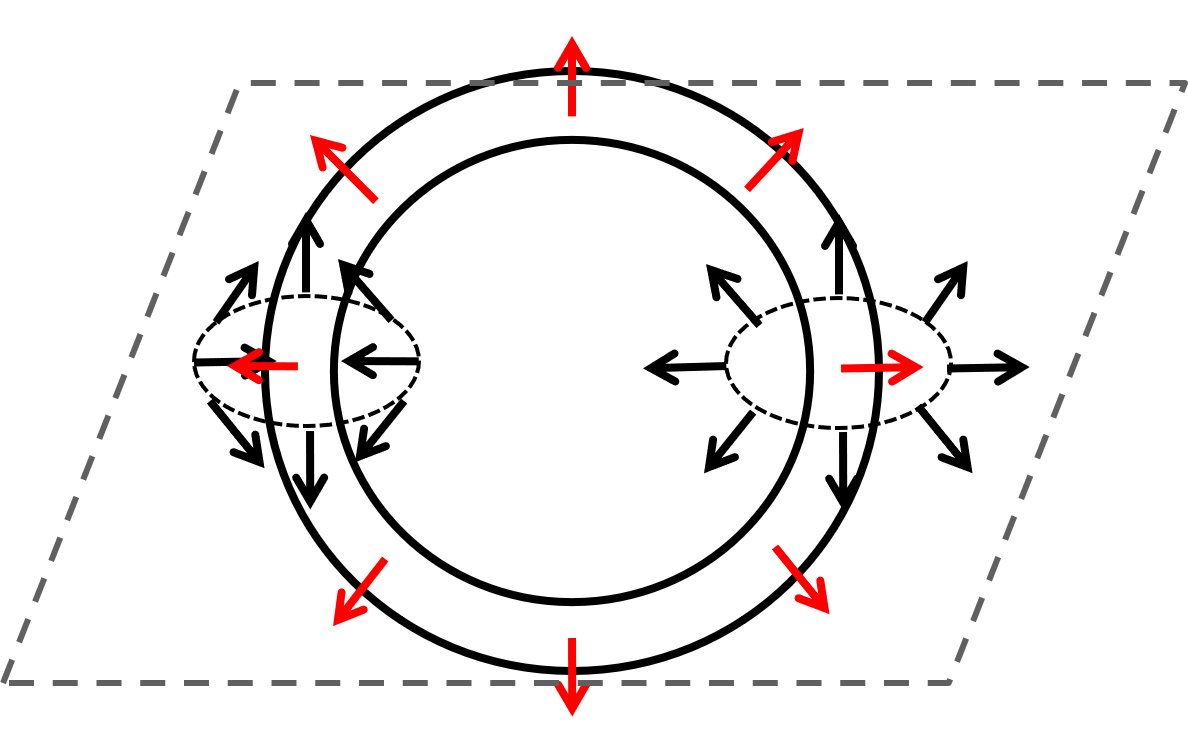}
&
\includegraphics[totalheight=4.5cm]{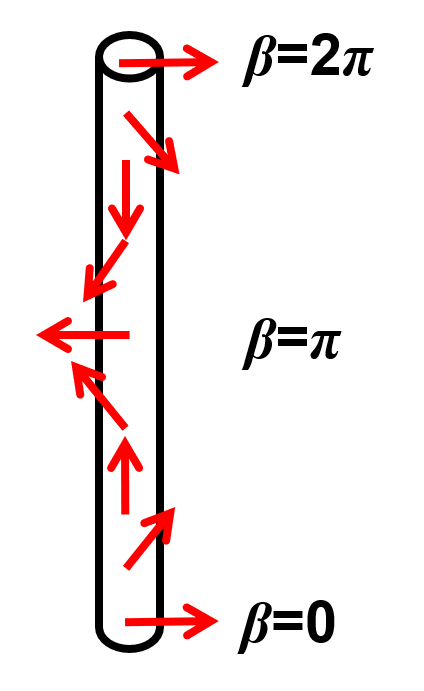}\\
(a) & (b)
\end{tabular}
\caption{A monopoles as a twisted Alice string.
(a) an isolated monopole as a twisted Alice ring 
along which the $U(1)$ modulus is twisted once.
(b) a monopole on a straight Alice string along which the $U(1)$ modulus is twisted once.
\label{fig:Alice-monopole}
}
\end{figure}

  An Alice string carries a $U(1)$ modulus
  $\beta$ in Eq.~(\ref{eq:general-Alice}), 
  corresponding to the internal direction of the flux 
  \cite{Chatterjee:2017hya}.
 Let us consider a closed Alice string as in Fig.~\ref{fig:Alice-monopole} (a).
 The black arrows represent the phase winding of the string. 
 The antipodal points of the ring (cut by a plane in the figure) 
 are a pair of a vortex and anti-vortex, 
 and therefore the ring is usually unstable to shrink.  
   When a $U(1)$ modulus $\beta$ (represented by red arrows) is twisted along the closed Alice string 
   as in Fig.~\ref{fig:Alice-monopole} (a), 
   it can be stable depending on a parameter region 
   and is called a vorton, as discussed below.
  This twisted Alice ring carries a magnetic monopole charge.

  In Ref.~\cite{Ruostekoski:2003qx}, 
Ruostekoski and Anglin found such a stable magnetic monopole 
solution in the theory without gauge fields 
(in the context of spinor Bose-Einstein condensates),
which is therefore a global monopole. 

A similar case for local monopoles were studied 
in the $SO(3)$ gauge theory with a fiveplet real scalar field 
(real symmetric traceless tensor) 
  \cite{Shankar:1976un,Bais:2002ae,Striet:2003na} admitting a conventional Alice string,
  in which the vacuum manifold is ${\mathbb R}P^2 \simeq S^2/{\mathbb Z}_2$ 
  rather than Eq.~(\ref{eq:OPS}).
    In this case, it was observed that a usual spherical `t Hooft-Polyakov monopole decays into a twisted Alice ring (in a certain parameter region).

Let us discuss the issue of the stability.
On the plane cutting the ring, the vortex and anti-vortex are given by
Eqs.~(\ref{eq:Alice-asymptotic}) and (\ref{eq:Alice-asymptotic-anti}), respectively. 
Note that because of the $U(1)$ modulus twist, 
the anti-vortex is not the one in Eq.~(\ref{eq:Alice-asymptotic2-anti}).
The $SU(2)$ fluxes are the same, while the $U(1)$ fluxes are opposite to each other.
In general, 
the $U(1)$ gauge field mediates an attraction between a vortex and anti-vortex, 
while the $SU(2)$ gauge fields mediates a repulsion between the same magnetic fluxes.
Thus, at least when the $SU(2)$ gauge field is lighter than 
the $U(1)$ gauge field ($g<e$)
and in the type-II regime (implying that 
Higgs fields are heavier than gauge fields), 
they can repel and contribute to the stability.

By cutting the Alice ring and stretch it to a straight string 
(along the $z$-direction), 
we obtain an Alice string along which 
the $U(1)$ modulus is twisted once, 
as in Fig.~\ref{fig:Alice-monopole} (b):
  \begin{eqnarray}
&& \Phi(r, \theta)  \sim 
\xi e^{i\frac{\theta}{2}} e^{+i\frac{\theta}{4}\hat \sigma} \sigma^1 e^{-i\frac{\theta}{4}\hat\sigma}, \nonumber 
\\
&& A_i \sim - \frac{1}{4 g}\frac{\epsilon_{ij} x^j}{r^2} \hat \sigma, \quad 
a_i \sim - \frac{1}{2 e_{}}\frac{\epsilon_{ij} x^j}{r^2},
  \nonumber\\
  && \hat\sigma = \sin \beta(z) \sigma^2 + \cos \beta(z) \sigma^3.
  \label{eq:monopole-on-string}
\end{eqnarray}
In this case, the monopole charge density is broadly present  
where the $U(1)$ modulus is twisted. Namely, the $U(1)$ modulus $\beta$ of the string winds once along it from $\beta =0$ at $z \to - \infty$ to $\beta =0$ at 
$z \to + \infty$. 
However, this charge cannot be localized and is spread over the string 
in order to reduce the gradient energy.\footnote{
If one direction of the space is compactified on $S^1$, 
and the string winds around it, the configuration is stable.}


\section{Confinement of Alice strings and monopoles}\label{sec:confinement}

In this section, we first discuss the AB phase
 around an Alice string.
 We then introduce AB defects appearing when the doublet field  develops VEVs, and discuss the bulk-soliton moduli locking 
 in the deconfined phase. 
 After classifying configurations of two strings by AB phases of the doublet field,
 the confinement of Alice strings and decays of an ANO string and ${\mathbb Z}_2$ string are discussed in the confined phase.
 Finally, we discuss the confinement of monopoles.
 
\subsection{Aharonov-Bohm phases around an Alice string}
When the doublet field $\Psi = (a,b)^T$ encircles around an Alice string, 
it receives an AB phase as
\begin{eqnarray}
\Psi_\theta &=& e^{i \frac{e_{}}{2} \int_0^\theta a\cdot dl} \l(Pe^{i g \int_0^\theta A\cdot dl} \r)
\left(
\begin{array}{c}
  a  \\
  b 
\end{array}
\right) .
\end{eqnarray}
This phase depends on the $U(1)$ modulus of the Alice string.
For instance, it is given by
\begin{eqnarray}
\Psi_\theta
=  
\left\{
\begin{array}{c}
e^{i \frac{\theta}{4}} e^{+ i \frac{\theta}{4}\sigma^3}
 \left(
\begin{array}{c}
  a  \\
  b 
\end{array}
\right)
=
\left(
\begin{array}{c}
e^{i \frac{\theta}{2}}  a  \\
b
\end{array}
\right) 
\mbox{ for }
 \Phi
\sim 
 \xi \left(
\begin{array}{ccc}
  0 &  e^{ i \theta}     \\
  1 & 0  
\end{array}
\right)
\\
e^{i \frac{\theta}{4}} e^{- i \frac{\theta}{4}\sigma^3}
 \left(
\begin{array}{c}
  a  \\
  b 
\end{array}
\right)
=
 \left(
\begin{array}{c}
 a  \\
 e^{i \frac{\theta}{2}}  b
\end{array}
\right) 
\mbox{ for }
 \Phi
\sim 
 \xi \left(
\begin{array}{ccc}
                0 & 1     \\
  e^{ i \theta} & 0  
\end{array}
\right) 
\\
\begin{array}{c}
e^{-i \frac{\theta}{4}} e^{- i \frac{\theta}{4}\sigma^3}
 \left(
\begin{array}{c}
  a  \\
  b 
\end{array}
\right)
=
\left(
\begin{array}{c}
e^{-i \frac{\theta}{2}}  a  \\
b
\end{array}
\right) 
\mbox{ for }
 \Phi
\sim 
 \xi \left(
\begin{array}{ccc}
  0 &  e^{- i \theta}     \\
  1 & 0  
\end{array}
\right)
\\
e^{-i \frac{\theta}{4}} e^{+ i \frac{\theta}{4}\sigma^3}
 \left(
\begin{array}{c}
  a  \\
  b 
\end{array}
\right)
=
 \left(
\begin{array}{c}
 a  \\
 e^{-i \frac{\theta}{2}}  b 
\end{array}
\right) 
\mbox{ for }
 \Phi
\sim 
 \xi \left(
\begin{array}{ccc}
                 0 & 1     \\
  e^{-i \theta} & 0  
\end{array}
\right) 
\end{array}
\end{array}
\right.
.
\end{eqnarray}
As it can be noticed that after a complete encirclement, the doublet field gets an AB phase as
\begin{eqnarray}
\label{thetadoublet1}
\Psi_{2\pi} &=& e^{i \frac{e_{}}{2} \oint a\cdot dl} \l(Pe^{i g \oint A\cdot dl} \r)
\left(
\begin{array}{c}
 a  \\
 b 
\end{array}
\right) 
= 
\left\{\begin{array}{c}
\left(
\begin{array}{c}
 -  a  \\
 +  b 
\end{array}
\right) 
\mbox{ for }
 \Phi
\sim 
 \xi \left(
\begin{array}{ccc}
  0 &  e^{\pm i \theta}     \\
  1 & 0  
\end{array}
\right)
\\
\left(
\begin{array}{c}
 + a  \\
 - b 
\end{array}
\right) 
\mbox{ for }
 \Phi
\sim 
 \xi \left(
\begin{array}{ccc}
                0 & 1     \\
  e^{\pm i \theta} & 0  
\end{array}
\right) 
\end{array}
\right.  .
\end{eqnarray}

The important observation is that 
the doublet field can feel the flux (modulus) of 
the Alice string from infinitely far away from it 
by encircling around it.
From this fact, we say that color is not confined for a single Alice string.

\subsection{Bulk-soliton moduli locking}
When the doublet field $\Psi$ develops VEVs in the absence of an Alice string,
the VEV can be arbitrary:
\begin{eqnarray}
 \Psi = \left(
\begin{array}{c}
  a  \\
  b
\end{array}
\right)     , \quad |a|^2 + |b|^2 = v^2,
\end{eqnarray}
parametrizing $S^3$ (if we turn off the interaction 
between the triplet and doublet fields).

On the other hand,
in the presence of a single Alice string, 
the VEVs of $\Psi$ is single-valued for 
\begin{eqnarray}
 \left\{\begin{array}{c}
  \Psi = 
 \left(
\begin{array}{c}
  0  \\
  v
\end{array}
\right) ,
 \quad K 
 = (\mathbb{Z}_4)_{Y+ W_{3}}          
 \quad \mbox{ for }
 \Phi
\sim 
 \xi \left(
\begin{array}{ccc}
  0 &  e^{\pm i \theta}     \\
  1 & 0  
\end{array}
\right)  \\
 \Psi = 
 \left(
\begin{array}{c}
  v \\
  0
\end{array}
\right) ,
 \quad K 
 = (\mathbb{Z}_4)_{Y- W_{3}}
 \quad
\mbox{ for }
 \Phi
\sim 
 \xi \left(
\begin{array}{ccc}
                      0 & 1     \\
  e^{\pm i \theta} & 0  
\end{array}
\right)    
\end{array}
\right.
.
\label{eq:no-AB-defect}
\end{eqnarray}
More generally, the single-valuedness condition is given by 
\begin{eqnarray}
   \Psi = 
U \left(
\begin{array}{c}
  0  \\
  v
\end{array}
\right) 
=
 \left(
\begin{array}{c}
    i v \sin \frac{\beta}{2}      \\
    v \cos \frac{\beta}{2}
\end{array}
\right),
 \quad K = ({\mathbb Z}_4)_{Y+W_{2,3}} 
 \quad \mbox{ for }
\Phi \sim
 \xi U \left(
\begin{array}{ccc}
  0 &  e^{\pm i \theta}     \\
  1 & 0  
\end{array}
\right) U^\dagger 
\label{eq:no-AB-defect-general}
\end{eqnarray}
with $U$ defined in Eq.~(\ref{eq:U-sigma1}),  
for the (anti-)Alice string configurations in 
Eq.~(\ref{eq:general-Alice}) [(\ref{eq:Alice-asymptotic3-anti})].
Here, $\beta =0$ and $\beta=\pi$ 
correspond to 
the first and second lines in Eq.~(\ref{eq:no-AB-defect}), 
respectively (up to the constant factor $i$). 
Otherwise, the doublet field cannot be single-valued.
Thus, we have found that 
the $U(1)$ modulus of the Alice string and 
the vacuum moduli of the doublet fields are locked.
We call this the {\it bulk-soliton moduli locking mechanism}. 
We should point out that $\Psi$ is in the deconfined phase 
satisfies $\Psi^\dagger \Phi \Psi =0$  
as can be seen in Eq.~(\ref{eq:A-B2}).

We have two schemes to interpret this phenomenon. 
When the triplet VEVs are much larger than the doublet VEVs, 
the symmetry breaking by the triplet field 
occurs at high energy 
and subsequently the symmetry breaking by the doublet field occurs at low energy. 
This is the scheme that we are mainly considering in this paper.
In this case, we can fix a triplet configuration (with a string) 
as a background.
Then, if we turn on the doublet VEVs gradually, 
they develop as in Eq.~(\ref{eq:no-AB-defect}) 
or more generally as in Eq.~(\ref{eq:no-AB-defect-general}). 
Thus, the doublet choses the deconfined phase. 
It is interesting that the phase of matter is determined by 
the presence of a soliton.

In the opposite scheme discussed in Sec.~\ref{sec:opposite},
in which the doublet VEVs are much larger than the triplet VEVs, 
the symmetry breaking by the doublet field 
occurs at high energy 
and subsequently the symmetry breaking by the triplet field occurs at low energy.
In this case, if one creates an Alice string, 
its $U(1)$ modulus $\beta$ should be aligned to the doublet field, 
as Eq.~(\ref{eq:no-AB-defect}) 
or more generally as Eq.~(\ref{eq:no-AB-defect-general}). 
Namely, the flux is determined by the doublet moduli in the bulk.
This can be understood from the vacuum manifold in 
the first line in Eq.~(\ref{eq:opposite-vacuum-mfd})  allowing strings.
In this case, one can observe that strings are purely 
ANO-like, 
in contrast to Eq.~(\ref{eq:OPS}) 
in which the $U(1)$ group is present in the denominator of the coset space, endowing the $U(1)$ modulus to the string.
When one creates multiple Alice strings, 
 all of their $U(1)$ moduli must be aligned to the doublet field.
In other words, the $U(1)$ moduli of strings are killed by the doublet field.

\subsection{Aharonov-Bohm defects}

\begin{figure}[!htb]
\centering
\begin{tabular}{cc}
\includegraphics[totalheight=2.5cm]{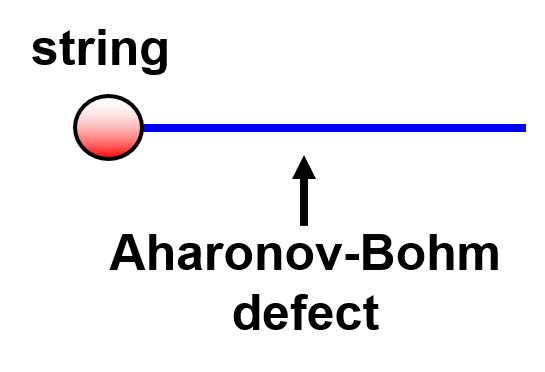}
&
\includegraphics[totalheight=2.5cm]{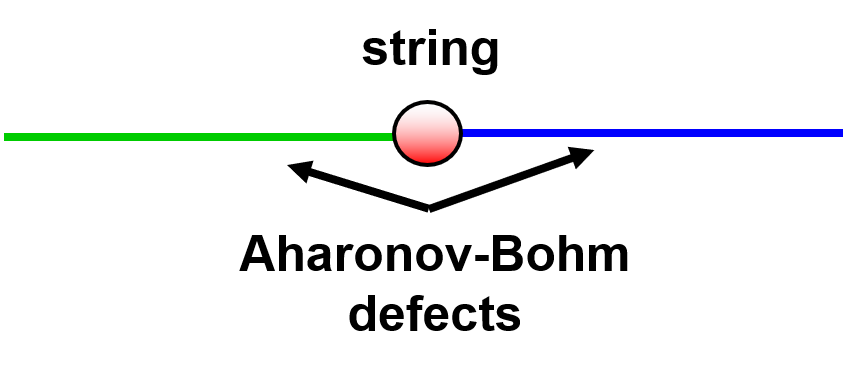}\\
(a) & (b)
\end{tabular}
\caption{Aharonov-Bohm defects attached to a string. 
(a) one or (b) two AB defect(s) attached 
to cancel the nontrivial AB phase(s) of the doublet fields around the string.
(a) is the case of an Alice string.
(b) corresponds to the case of an ANO string or ${\mathbb Z}_2$ string.
The two AB defects are generated in the two components of the doublet field, 
and in general they have different tension depending on the VEV of each component.
\label{fig:AB}
}
\end{figure}

Next, let us consider the confined phase.\footnote{
Or, we enforce the doublet to have a VEV 
on the opposite side even in the deconfined phase.
}
Then, 
it is inevitable for the doublet fields to receive an AB phase 
which must be canceled for the single-valuedness as
\begin{eqnarray}
 \Psi \sim  
 \left\{\begin{array}{c}
 \left(
\begin{array}{c}
    a  e^{\pm i \frac{\theta}{2}}  f(\pm (\theta-\alpha))  \\
    b
\end{array}
\right)      
\mbox{ for }
 \Phi
\sim 
 \xi \left(
\begin{array}{ccc}
  0 &  e^{\pm i \theta}     \\
  1 & 0  
\end{array}
\right)  \\
 \left(
\begin{array}{c}
  a \\
  b  e^{\pm i \frac{\theta}{2}}  f(\pm(\theta-\alpha))
\end{array}
\right)      
\mbox{ for }
 \Phi
\sim 
 \xi \left(
\begin{array}{ccc}
                0 & 1     \\
  e^{\pm i \theta} & 0  
\end{array}
\right) 
\end{array}
\right.
,
\label{eq:AB-defect}
\end{eqnarray}
where the factors $e^{\pm i \frac{\theta}{2}}$ come from the AB-phase,
and $\alpha$ corresponds to the direction of the AB defect.
Here, for the single-valuedness of $\Psi$, we have introduced a multi-valued function  $f(\theta)$  on 
$S^1$ which changes the sign under a $2\pi$ shift:
$f(\theta + 2\pi) = - f(\theta)$.
This creates a soliton attached to the Alice string as schematically shown in Fig.~\ref{fig:AB}(a).
The existence of the soliton is topologically supported by 
the nontrivial zeroth homotopy group in Eq.~(\ref{eq:homotopy-doublet}) in the confined phase.
We called such a soliton as an AB defect \cite{Chatterjee:2019zwx}. 

There are two ways to create an AB defect.
\begin{enumerate}
\item
An ansatz for the doublet for the first case is to take
$f(\theta) = e^{i \phi(\theta)/2}$
 \cite{Chatterjee:2019zwx}: 
\begin{eqnarray}
\Psi_{\rm DW} \sim \xi_\psi \left(
\begin{array}{c}
 e^{i \l(\frac{\theta + \phi(\theta)}2  \r)}  \\
0
\end{array}
\right),\,\, \phi(0) = 0,\,\, \phi(2\pi) = -2\pi ,\label{eq:doublet-ansatz1}
\end{eqnarray}
where $\phi(\theta)$ is a decreasing function and 
the boundary condition 
keeps the doublet single-valued.
We substitute this ansatz 
 to the  Hamiltonian density in the presence of the Alice string configuration 
 in Eq.~(\ref{eq:Alice-asymptotic})
 at large distances, 
where the potentials are given in Eqs.~(\ref{eq:VPsi}) and (\ref{eq:int1}). 
We thus obtain the effective Hamiltonian of the doublet
\begin{eqnarray}
\mathcal{H}_{\rm DW}/\xi_{\psi}^2 \sim \frac{1}{4}\l[(\p_i\phi)^2 + 8 \mu_\psi^2 \l(1 - \cos\phi\r)\r], 
\quad \mu_\psi^2 = \xi\mu,
\end{eqnarray}
which is nothing but 
the sine-Gordon model. 

\item
The second case is to consider a kink configuration by a real function $f$ 
varying from $+1$ to $-1$ at $\theta=0$ to $\theta = 2\pi$.

\end{enumerate}

In either case, 
the AB defect is a linearly extended to infinity, 
with having infinite energy in infinite space.
So it should decay and 
the configurations in Eq.~(\ref{eq:AB-defect}) 
go to those in Eq.~(\ref{eq:no-AB-defect}) 
by changing the orientation of $\Psi$ in the internal space, 
and the phase is in the deconfined phase.
We conclude that the bulk-soliton moduli locking mechanism 
is energetically realized.

\bigskip
On the other hand, there appear two AB defects 
around an ANO string and a ${\mathbb Z}_2$ string 
as in Fig.~\ref{fig:AB}(b), 
since the both components of the doublet field 
$\Psi$ receive AB phases around those strings. 
The doublet fields behave around an ANO string as 
\begin{eqnarray}
 \Psi \sim  
 \left(
\begin{array}{c}
    a  e^{i \frac{\theta}{2}}  f_1(\theta)  \\
    b  e^{i \frac{\theta}{2}} f_2(\theta - \alpha) 
\end{array}
\right)    
\mbox{ for } \Phi (r, \theta)  
\sim 
 \xi \left(
\begin{array}{ccc}
  0 &  e^{ i \theta}     \\
    e^{ i \theta}  & 0  
\end{array}
\right) \label{eq:ANO-AB}
\end{eqnarray}
where $f_1$ and $f_2$ are multi-valued functions, 
$\alpha$ controls the relative directions of the AB defect 
(we have taken a constant in the upper component to be zero 
without loss of generality). 
Energetically, it may be natural to consider $\alpha = \pi$ as in 
Fig.~\ref{fig:AB}(b).
The two different AB defects are present in the two different components
of the doublet field $\Psi$, and so they have different tensions in general, 
depending on $a$ and $b$. 

Similarly,
the doublet fields behave around a
${\mathbb Z}_2$ string as
\begin{eqnarray}
 \Psi \sim  
 \left(
\begin{array}{c}
    a  e^{i \frac{\theta}{2}}  f_1(\theta)  \\
    b  e^{-i \frac{\theta}{2}} f_2(- \theta + \alpha) 
\end{array}
\right)    
\mbox{ for } \Phi (r, \theta)  
\sim 
 \xi \left(
\begin{array}{ccc}
                     0 &  e^{ i \theta}     \\
   e^{ - i \theta}  & 0  
\end{array}
\right) \label{eq:Z2-AB}
\end{eqnarray}
with multi-valued functions $f_1$ and $f_2$.

In these cases, it is impossible to eliminate the two AB defects 
simultaneously in any choice of the doublet VEVs,
although one can eliminate at least one AB defect 
by taking either $a=0$ or $b=0$.

\bigskip
In the case of $q=0$, one cannot eliminate an AB defect.

\subsection{Classification of configurations of two strings}
\begin{figure}[!htb]
\centering

\begin{tabular}{cc}
\includegraphics[totalheight=5.2cm]{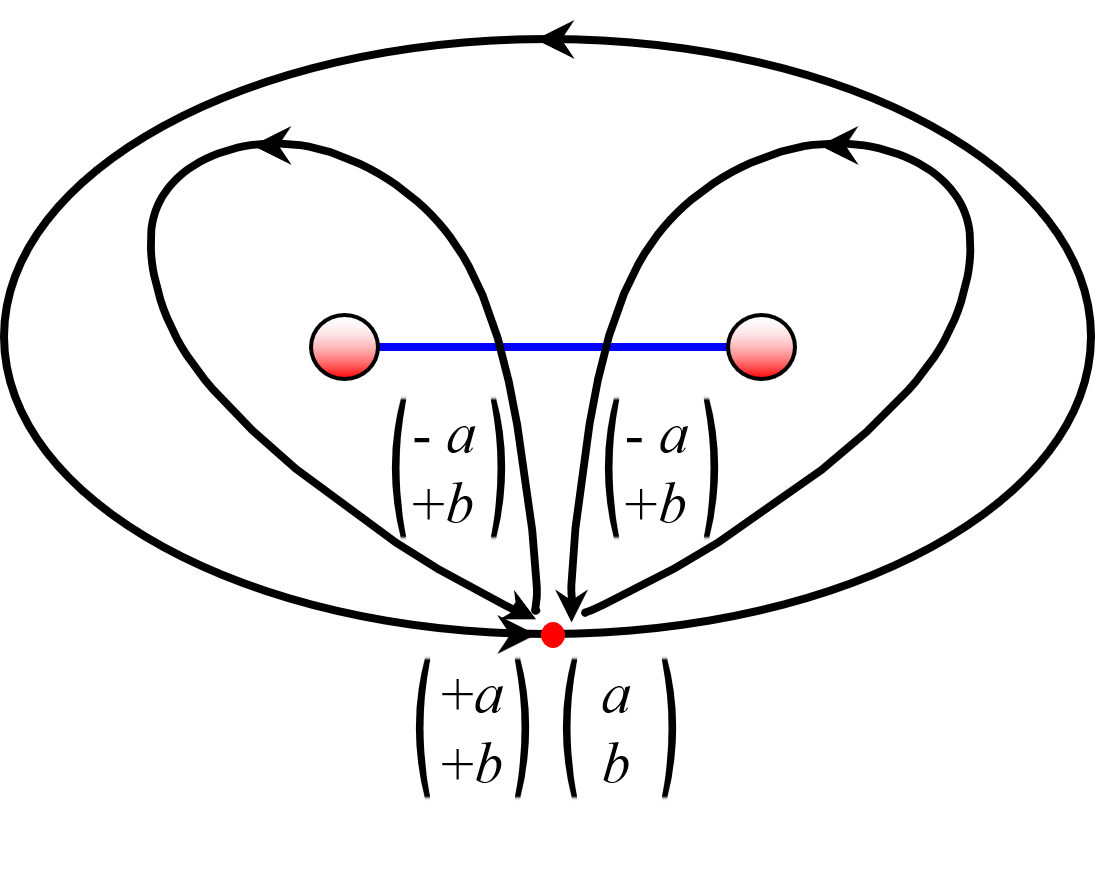}
&
\includegraphics[totalheight=5.2cm]{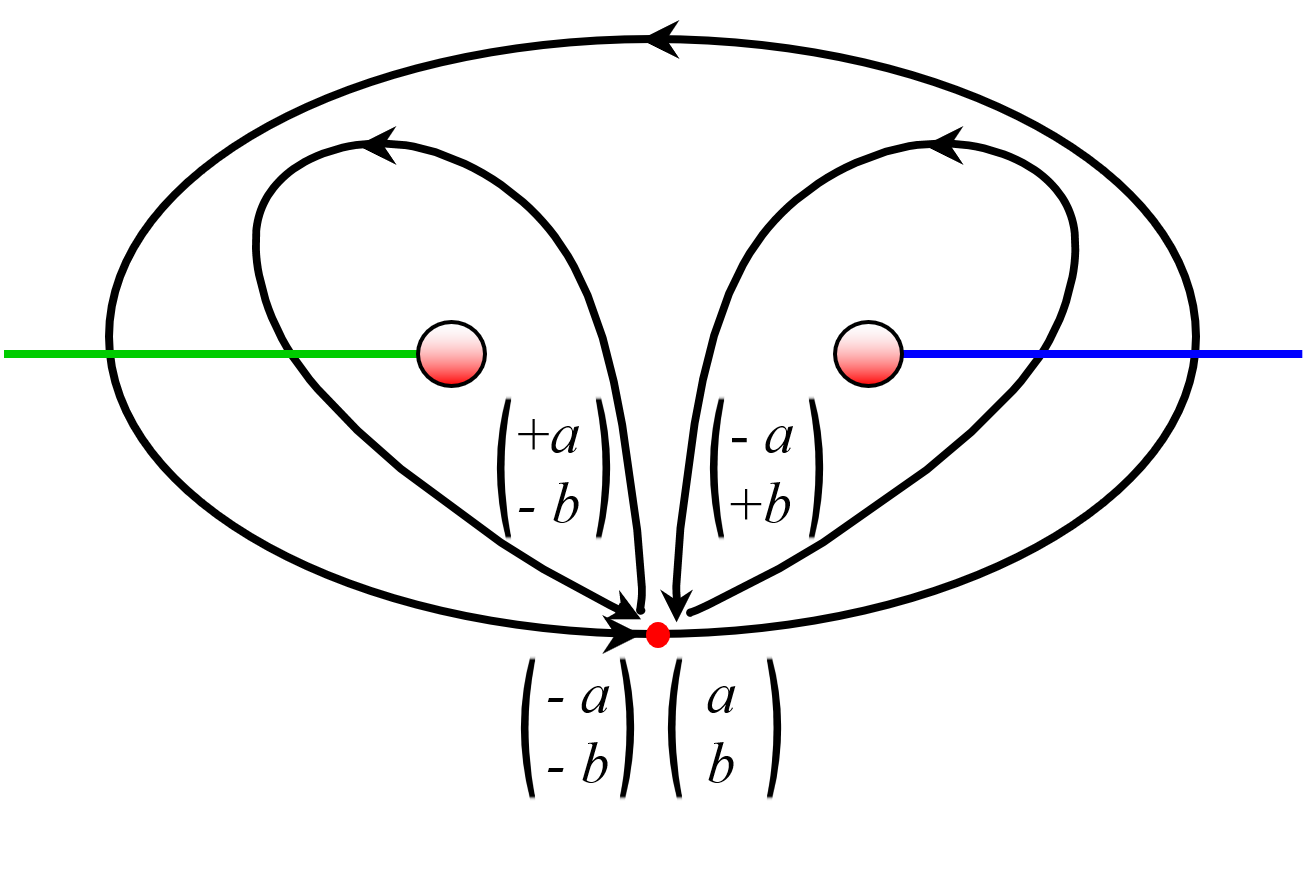}\\
(a) & (b)
\end{tabular}
\caption{Configurations of two strings classified by 
AB phases around the strings.
Configurations of two strings can be classified into 
either (a) the confinement or (b) the decay, depending on 
AB phases of the doublet field encircling around the strings.
(a) 
When the doublet field $\Psi =(a,b)^T$ encircles the right (left) string 
along a smaller loop, 
  it becomes $\Psi =(-a,b)^T$ which is not single-valued 
  if there is no AB defect, 
  implying the existence of an AB defect attached to each of them.
  When the doublet field $\Psi =(a,b)^T$ encircles the both strings together (along the large loop), 
  it comes back to $\Psi =(a,b)^T$ which is single-valued,
  implying the absence of AB defect crossing the large loop. 
  Therefore, one AB defect connects the two strings, corresponding to the confinement.
  (b) 
  When the doublet field $\Psi =(a,b)^T$ encircles the right (left) string, 
  it becomes $\Psi =(-a,b)^T$ ($\Psi =(a,-b)^T$) which is not single-valued if there is no AB defect, 
  implying the existence of a {\it different} AB defect attached to each of them.
  When the doublet field $\Psi =(a,b)^T$ encircles the both strings together (along the large loop), 
  it becomes $\Psi =(-a,-b)^T$ which is not yet single-valued,
  implying the existence of the two kinds of 
  AB defects crossing the large loop. 
  No AB defect connects the two strings, corresponding to the decay.
\label{fig:AB-phase}
}
\end{figure}

When the double VEVs are much larger than 
the triplet VEVs, we can fix a doublet configuration as a background. 
Let us consider the confined phase in which 
the both components of the doublet are nonzero 
and the ${\mathbb Z}_4$ symmetry is broken to ${\mathbb Z}_2$.
In this case, Alice strings are inevitably attached by AB defects. 
Here we discuss that configurations of two strings can be 
classified by the AB phases of the doublet field 
encircling around the two strings,  
as in Fig.~\ref{fig:AB-phase}.

The configurations of two strings can be classified into 
either  the confinement [Fig.~\ref{fig:AB-phase}(a)] 
or the decay [Fig.~\ref{fig:AB-phase}(b)], depending on 
AB phases of the doublet field encircling around the strings. 
Without loss of generality,  let us consider the case that 
when the doublet field $\Psi =(a,b)^T$ encircles the right string 
along a smaller loop, 
  it becomes $\Psi =(-a,b)^T$ which is not single-valued 
  if there is no AB defect, 
  implying the existence of an AB defect attached to it.   
Next,  when the doublet field $\Psi =(a,b)^T$ encircles the left string, 
let us suppose that it becomes either 
(a) $\Psi =(-a,b)^T$ or (b) $\Psi =(a,-b)^T$. 
In either case, an AB defect of a different type 
is attached to the left string.

  In the case of (a), the two AB defects attached to the two strings are of the same type, and so they can be connected. 
  In fact, if  the doublet field $\Psi =(a,b)^T$ encircles the both strings together along the larger loop in  Fig.~\ref{fig:AB-phase}(a), 
  it comes back to $\Psi =(a,b)^T$ which is single-valued.
  This shows the absence of AB defect crossing the large loop, 
  implying that the two strings must be connected by a single AB defect. This case implies the confinement of the two strings.
  
  On the other hand,
in the case of (b), the two AB defects attached to the two strings are of the different type, and so they cannot be connected. 
In fact,  if the doublet field $\Psi =(a,b)^T$ encircles the both strings together along the larger loop in  Fig.~\ref{fig:AB-phase}(b), 
  it becomes $\Psi =(-a,-b)^T$ which is not yet single-valued,
  implying the existence of the two kinds of 
  AB defects crossing the large loop.
  Thus,  no AB defect connects the two strings, 
  implying that the two strings are pulled to the opposite (or different) directions. 
  
  In the following subsections, we discuss the confinement
   in Fig.~\ref{fig:AB-phase}(a) 
   and the decay in Fig.~\ref{fig:AB-phase}(b),  respectively.

\subsection{Confinement of Alice strings}
First, let us discuss the case of confinement 
corresponding to Fig.~\ref{fig:AB-phase}(a).
There are two types of the confinement:
1) a pair of an Alice string and anti-Alice string (colorless confinement) and 
2) a pair of the same type of Alice strings (colorful confinement).
Let us discuss them one by one.

\begin{figure}[!htb]
\centering
\includegraphics[totalheight=3.5cm]{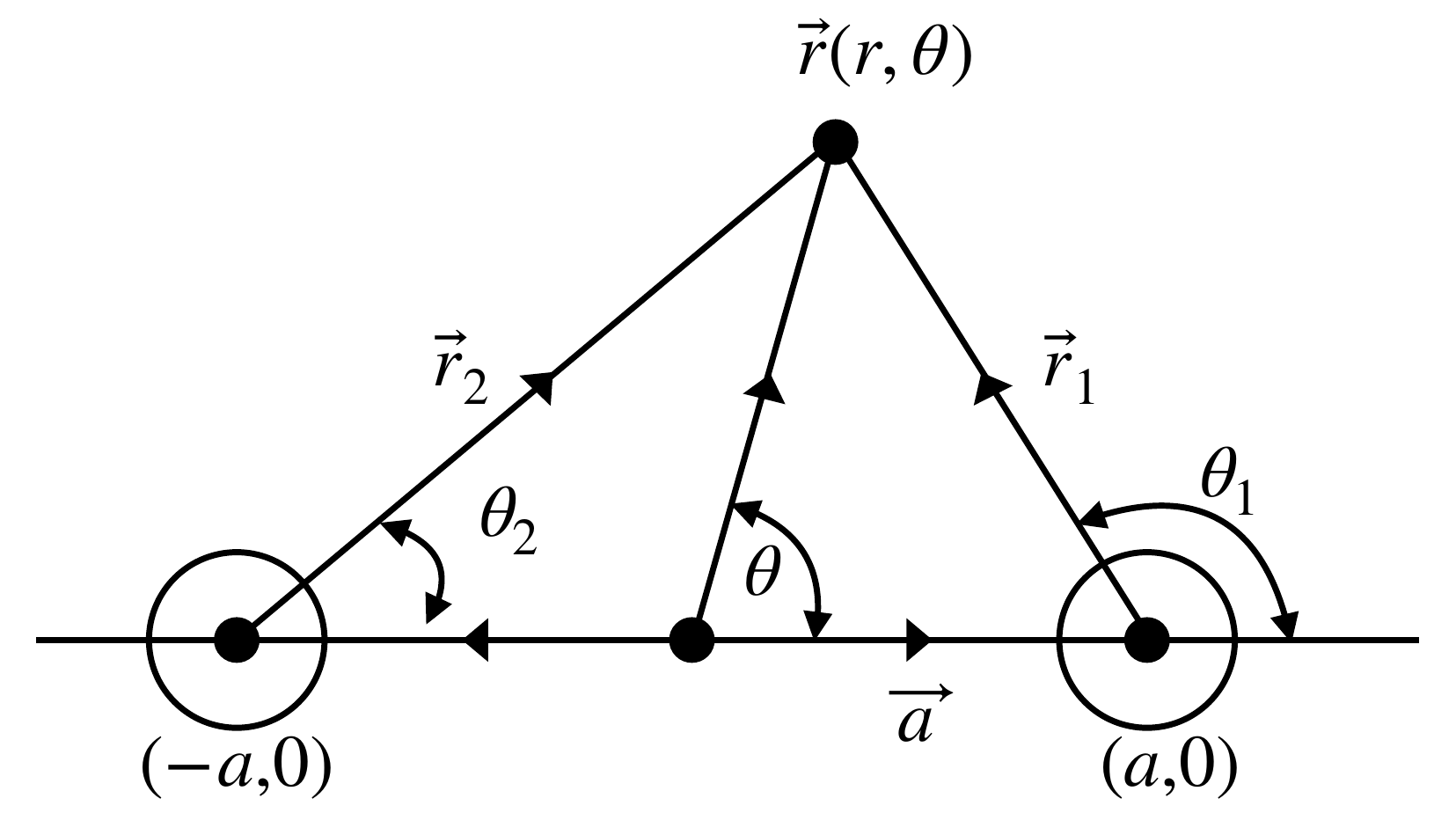} 
\caption{Configuration of two strings. 
The right (left) strings are placed at $(a,0)$ [$(-a,0)$].
$(r,\theta)$ are polar coordinates from the origin, 
and $(r_1,\theta_1)$ [$(r_2,\theta_2)$] are polar coordinates
 from $(a,0)$ [$(-a,0)$].
\label{fig:two-vortices}
}
\end{figure}

1) A pair of an Alice string and anti-Alice string (colorless confinement).\\
Let us consider a pair of configurations 
\begin{eqnarray}
&& \Phi_{1}(r, \theta)  
\sim 
 \xi \left(
\begin{array}{ccc}
  0 &  e^{ i \theta_1}     \\
  1 & 0  
\end{array}
\right) 
= \xi e^{i\frac{\theta_1}{2}} e^{+i\frac{\theta_1}{4}\sigma^3} \sigma^1 e^{-i\frac{\theta_1}{4}\sigma^3}, 
\nonumber  \\
&& 
A_{1,i} \sim - \frac{1}{4 g}\frac{\epsilon_{ij} x_1^j}{r_1^2} \sigma^3, \quad 
a_{1,i} \sim - \frac{1}{2 e_{}}\frac{\epsilon_{ij} x_1^j}{r_1^2}, \nonumber \\
&&
\Psi_1 \sim \left(
\begin{array}{c}
    a  e^{i \frac{\theta_1}{2}}  f_1(\theta_1- \alpha_1)  \\
    b 
\end{array}
\right)    
\end{eqnarray}
and 
\begin{eqnarray}
&& \Phi_{2}(r, \theta)  
\sim 
 \xi \left(
\begin{array}{ccc}
  0 &   e^{ -i \theta_2}    \\
  1 & 0  
\end{array}
\right) 
= \xi e^{-i\frac{\theta_2}{2}} e^{-i\frac{\theta_2}{4}\sigma^3} \sigma^1 e^{+i\frac{\theta_2}{4}\sigma^3}, \nonumber\\
&& 
A_{2,i} \sim + \frac{1}{4 g}\frac{\epsilon_{ij} x_2^j}{r_2^2} \sigma^3, \quad 
a_{2,i} \sim + \frac{1}{2 e_{}}\frac{\epsilon_{ij} x_2^j}{r_2^2},\nonumber \\
&&
\Psi_2 \sim \left(
\begin{array}{c}
    a  e^{-i \frac{\theta_2}{2}}  f_1(\theta_2- \alpha_2)  \\
    b 
\end{array}
\right)    
\end{eqnarray}
with constants  $\alpha_1$ and $\alpha_2$ describing the directions of the AB defects.
Here $(r,\theta)$ are polar coordinates from the origin, 
and $(r_1,\theta_1)$ [$(r_2,\theta_2)$] are polar coordinates
 from $(a,0)$ [$(-a,0)$].
The configurations $(\Phi_1, A_1, a_1,\Psi_1)$ and  $(\Phi_2, A_2, a_2,\Psi_2)$ 
denote a vortex placed at 
$(a,0)$ and anti-vortex at $(-a,0)$. 
These two vortices are connected by an AB defect as 
in Fig.~\ref{fig:AB-phase}(a),
if one considers the total configuration constructed by an Abrikosov-like ansatz,
\begin{eqnarray}
&& \Phi_{\rm tot}  \sim 
 \xi \left(
\begin{array}{ccc}
  0 &   e^{ i(\theta_1 - \theta_2)}    \\
  1 & 0  
\end{array}
\right)  \to  
 \xi \left(
\begin{array}{ccc}
  0 & 1   \\
  1 & 0  
\end{array}
\right) ,\nonumber\\
&& A_{\rm tot} = A_{1} + A_{2} \to 0, 
\quad a_{\rm tot} = a_{1} + a_{2} \to 0,   \nonumber \\
&& \Psi_{\rm tot} \sim  
 \left(
\begin{array}{c}
    a  e^{i \frac{\theta_1-\theta_2}{2}}  f_1(\theta_1-\theta_2-\pi)  \\
    b 
\end{array}
\right)    
 \to 
 \left(
\begin{array}{c}
    a \\
    b 
\end{array}
\right)    .
\end{eqnarray}
This configuration corresponds to take 
$\alpha_1=\pi, \alpha_2=0$ in the individual configurations in order to 
connect them.
In this case, they annihilate in pair as indicated by arrows in the above equation, and the final state is
the vacuum $\Phi = \sigma^1$ and $\Psi = {\rm constant}$ as expected.

\bigskip

2) A pair of the same type of Alice strings (colorful confinement).\\
Next, let us consider a pair of configurations 
\begin{eqnarray}
&& \Phi_{1}(r, \theta)  
\sim 
 \xi \left(
\begin{array}{ccc}
  0 &  e^{ i \theta_1}     \\
  1 & 0  
\end{array}
\right) 
= \xi e^{i\frac{\theta_1}{2}} e^{+i\frac{\theta_1}{4}\sigma^3} \sigma^1 e^{-i\frac{\theta_1}{4}\sigma^3}, 
\nonumber  \\
&& 
A_{1,i} \sim - \frac{1}{4 g}\frac{\epsilon_{ij} x_1^j}{r_1^2} \sigma^3, \quad 
a_{1,i} \sim - \frac{1}{2 e_{}}\frac{\epsilon_{ij} x_1^j}{r_1^2},\\
\nonumber  \\
&&
\Psi_1 \sim \left(
\begin{array}{c}
    a  e^{i \frac{\theta_1}{2}}  f_1(\theta_1- \alpha_1)  \\
    b 
\end{array}
\right)    
\end{eqnarray}
and 
\begin{eqnarray}
&& \Phi_{2}(r, \theta)  
\sim 
 \xi \left(
\begin{array}{ccc}
  0 &  e^{ i \theta_2}     \\
  1 & 0  
\end{array}
\right) 
= \xi e^{i\frac{\theta_2}{2}} e^{+i\frac{\theta_2}{4}\sigma^3} \sigma^1 e^{-i\frac{\theta_2}{4}\sigma^3}, 
\nonumber  \\
&& 
 A_{2,i} \sim - \frac{1}{4 g}\frac{\epsilon_{ij} x_2^j}{r_2^2} \sigma^3, \quad 
 a_{2,i} \sim - \frac{1}{2 e_{}}\frac{\epsilon_{ij} x_2^j}{r_2^2},\nonumber \\
&&
\Psi_2 \sim \left(
\begin{array}{c}
    a  e^{i \frac{\theta_2}{2}}  f_1(\theta_2- \alpha_2)  \\
    b 
\end{array}
\right)    .
\end{eqnarray}
The coordinates are the same with Fig.~\ref{fig:two-vortices},
and 
the configurations $(\Phi_1, A_1, a_1)$ and  $(\Phi_2, A_2, a_2)$ 
denote vortices of the same kind 
placed at $(a,0)$ and $(-a,0)$. 
Again, these two vortices are also connected by an AB defect as 
in Fig.~\ref{fig:AB-phase}(a),
if the total configuration is constructed by an Abrikosov-like ansatz,
\begin{eqnarray}
&& \Phi_{\rm tot}  \sim 
 \xi \left(
\begin{array}{ccc}
  0 &   e^{ i(\theta_1 + \theta_2)}    \\
  1 & 0  
\end{array}
\right)  \to  
 \xi \left(
\begin{array}{ccc}
  0 & e^{2i \theta}   \\
  1 & 0  
\end{array}
\right) ,\nonumber\\
&& A_{\rm tot} = A_{(1)} + A_{(2)} 
\to 
 - \frac{1}{2 g}\frac{\epsilon_{ij} x^j}{r^2} \sigma^3, \quad 
\quad 
 a_{\rm tot} = a_{(1)} + a_{(2)} \to
 - \frac{1}{ e_{}}\frac{\epsilon_{ij} x^j}{r^2}, \nonumber \\
&& \Psi_{\rm tot} \sim  
 \left(
\begin{array}{c}
    a  [e^{i \frac{\theta_1+\theta_2}{2}} +  
       e^{i \frac{\theta_1-\theta_2}{2}}  f_1(\theta_1-\theta_2-\pi) ] 
      \\
    b
\end{array}
\right)    
 \to 
 \left(
\begin{array}{c}
    a e^{i\theta}\\
    b 
\end{array}
\right)     .
\end{eqnarray}
Then, the final state indicated after the arrows will be the doubly-wound Alice string configuration 
with a singly quantized $U(1)$ magnetic flux and half-quantized $SU(2)$ magnetic 
flux in 
Eq.~(\ref{eq:Alice-asymptotic-double}). 
In this case, the $SU(2)$ color flux remains in the confined string.
One may think that color is not confined.
However, this color flux is screened and cannot be seen from the large distance by AB phases.

If we consider the deconfined phase, 
either of the two components of the doublet is zero, 
the ${\mathbb Z}_4$ symmetry is unbroken, 
and domain walls are not topologically supported.
In this case, either of ANO string is free from an AB defect
as can be seen in Eq.~(\ref{eq:no-AB-defect}), 
and consequently it is not confined (but the other is confined).

Comparing the vacuum manifolds for the deconfined and confined phases in 
Eq.~(\ref{eq:G/K}), one can understand that 
the confined phase does not admit a half $U(1)$ winding (and therefore 
a half-quantized $U(1)$ magnetic flux), 
while the deconfined phase does, 
consistent with the above discussion. 

From the vacuum manifolds
in Eq.~(\ref{eq:opposite-vacuum-mfd})
 for the second symmetry breaking 
in the opposite hierarchy, 
one can understand that 
in the confined phase strings are half-quantized strings 
corresponding to doubly-wound Alice strings,
in contrast to 
the deconfined phase admitting 1/4-quantized strings
corresponding to the Alice strings.

\subsection{Decay of Abrikosov-Nielsen-Olesen string and ${\mathbb Z}_2$ string}
Next, let us discuss the case of decay  
corresponding to Fig.~\ref{fig:AB-phase}(b).

An ANO vortex is attached by two AB defects in the confined phase  
as in Eq.~(\ref{eq:ANO-AB}) and Fig.~\ref{fig:AB} (b).
It decays and is split into a pair of Alice strings
with the opposite $SU(2)$ fluxes:
\begin{eqnarray}
&& \Phi_1(r, \theta)  
\sim 
 \xi \left(
\begin{array}{ccc}
  0 &  e^{ i \theta_1}     \\
  1 & 0  
\end{array}
\right) 
= \xi e^{i\frac{\theta_1}{2}} e^{+i\frac{\theta_1}{4}\sigma^3} \sigma^1 e^{-i\frac{\theta_1}{4}\sigma^3}, 
\nonumber  \\
&& 
A_{1,i} \sim - \frac{1}{4 g}\frac{\epsilon_{ij} x_1^j}{r_1^2} \sigma^3, \quad 
a_{1,i} \sim - \frac{1}{2 e_{}}\frac{\epsilon_{ij} x_1^j}{r_1^2},\\
\nonumber  \\
&&
\Psi_1 \sim \left(
\begin{array}{c}
    a  e^{i \frac{\theta_1}{2}}  f_1(\theta_1- \alpha_1)  \\
    b 
\end{array}
\right)    
\end{eqnarray}
and 
\begin{eqnarray}
&& \Phi_2(r, \theta)  
\sim  \xi \left(
\begin{array}{ccc}
                 0 &  1   \\
  e^{ i \theta_2}  & 0  
\end{array}
\right) 
= \xi e^{i\frac{\theta_2}{2}} e^{-i\frac{\theta_2}{4}\sigma^3} \sigma^1 e^{+i\frac{\theta_2}{4}\sigma^3},
\nonumber  \\
&& 
A_{2,i} \sim + \frac{1}{4 g}\frac{\epsilon_{ij} x_2^j}{r_2^2} \sigma^3, \quad 
a_{2,i} \sim - \frac{1}{2 e_{}}\frac{\epsilon_{ij} x_2^j}{r_2^2},
\nonumber  \\
&&
\Psi_2 \sim \left(
\begin{array}{c}
    a  \\
    b  e^{i \frac{\theta}{2}} f_2(\theta - \alpha_2) 
\end{array}
\right)    .
\end{eqnarray}
Here, the coordinates are the same with Fig.~\ref{fig:two-vortices},
and 
the configurations $(\Phi_1, A_1, a_1,\Psi_1)$ and  $(\Phi_2, A_2, a_2,\Psi_2)$ 
denote vortices 
placed at $(a,0)$ and $(-a,0)$ as in  
Fig.~\ref{fig:AB-phase}(b). 
These two vortices cannot be connected by an AB defect. Instead, 
two different AB defects are attached to each of them and 
pull the vortices to the opposite (or different) directions.

\bigskip
A ${\mathbb Z}_2$ vortex is also attached by two AB defects in the confined phase as in Eq.~(\ref{eq:Z2-AB}) and Fig.~\ref{fig:AB} (b). 
It also decays and is split into a pair of Alice string 
and anti-Alice string 
with the same $SU(2)$ magnetic fluxes:
\begin{eqnarray}
&& \Phi_1(r, \theta)  
\sim 
 \xi \left(
\begin{array}{ccc}
  0 &  e^{ i \theta_1}     \\
  1 & 0  
\end{array}
\right) 
= \xi e^{i\frac{\theta_1}{2}} e^{+i\frac{\theta_1}{4}\sigma^3} \sigma^1 e^{-i\frac{\theta_1}{4}\sigma^3}, 
\nonumber  \\
&& 
A_{1,i} \sim - \frac{1}{4 g}\frac{\epsilon_{ij} x_1^j}{r_1^2} \sigma^3, \quad 
a_{1,i} \sim - \frac{1}{2 e_{}}\frac{\epsilon_{ij} x_1^j}{r_1^2},
\nonumber  \\
&&
\Psi_1 \sim \left(
\begin{array}{c}
    a  e^{i \frac{\theta_1}{2}}  f_1(\theta_1- \alpha_1)  \\
    b 
\end{array}
\right)    
\end{eqnarray}
and 
\begin{eqnarray}
&& \Phi_2(r, \theta)  
\sim  \xi \left(
\begin{array}{ccc}
                 0 &  1   \\
  e^{- i \theta_2}  & 0  
\end{array}
\right) 
= \xi e^{-i\frac{\theta_2}{2}} e^{+i\frac{\theta_2}{4}\sigma^3} \sigma^1 e^{-i\frac{\theta_2}{4}\sigma^3},
\nonumber  \\
&& 
A_{2,i} \sim - \frac{1}{4 g}\frac{\epsilon_{ij} x_2^j}{r_2^2} \sigma^3, \quad 
a_{2,i} \sim + \frac{1}{2 e_{}}\frac{\epsilon_{ij} x_2^j}{r_2^2},
\nonumber  \\
&&
\Psi_2 \sim \left(
\begin{array}{c}
    a  \\
    b  e^{-i \frac{\theta_2}{2}} f_2(-\theta_2 + \alpha_2) 
\end{array}
\right)   .
\end{eqnarray}
Again, 
 the coordinates are the same with Fig.~\ref{fig:two-vortices},
and 
the configurations $(\Phi_1, A_1, a_1,\Psi_1)$ and  $(\Phi_2, A_2, a_2,\Psi_2)$ 
denote vortices 
placed at $(a,0)$ and $(-a,0)$ as in  
Fig.~\ref{fig:two-vortices}. 
These two vortices are attached and pulled by 
the two different AB defects as 
in Fig.~\ref{fig:AB-phase}(b).

If we consider the deconfined phase in which 
either of the two components of the doublet is zero,
only one AB defect is attached to the ANO (or ${\mathbb Z}_2$) string, 
and either of two Alice strings is free from an AB defect
as can be seen in Eq.~(\ref{eq:no-AB-defect}).
However, the decay of the ANO string and ${\mathbb Z}_2$ string 
always occurs since either of constituents is attached by 
an AB defect.

\subsection{Confinement of Alice monopoles: monopole mesons}

Here we turn to discuss that monopoles are also confined.
We first discuss the deconfined phase 
in which one of components of the doublet fields is zero:
$\Psi =(a,0)^T$.
Then, we point out that the same story holds for the confined phase as well.

First, let us consider a monopole on an Alice string.
Let an Alice string be placed along the $z$ direction 
as in Eq.~(\ref{eq:monopole-on-string}),
where  
and the $U(1)$ modulus $\beta$ of the string winds once along it from 
$\beta = 0$ at $z \to - \infty$ to $\beta =2\pi$ at 
$z \to + \infty$. 
Then, there is one monopole on the string at $\beta = \pi$ as 
Sec.~\ref{sec:monopole}.
In the absence of the doublet VEV, the monopole charge 
density is spread over the string. However, it is not the case 
in the presence of the doublet VEV.

\begin{figure}[!htb]
\centering
\begin{tabular}{cc}
\includegraphics[totalheight=5cm]{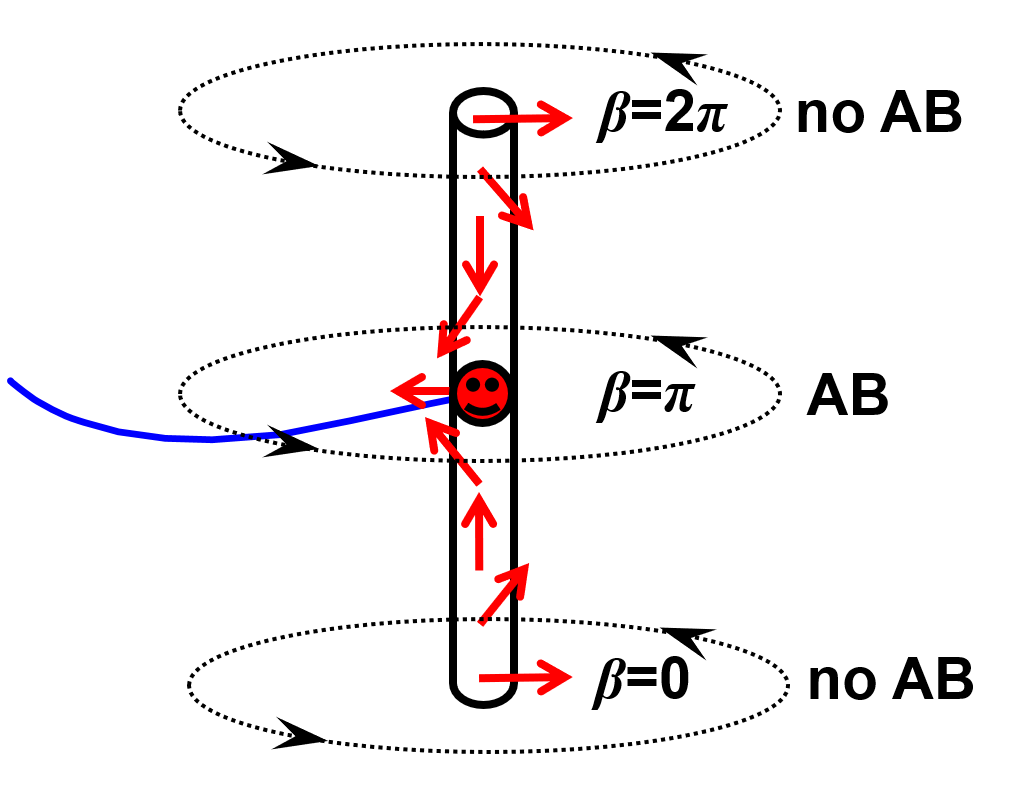} & \quad\quad\quad
\includegraphics[totalheight=5cm]{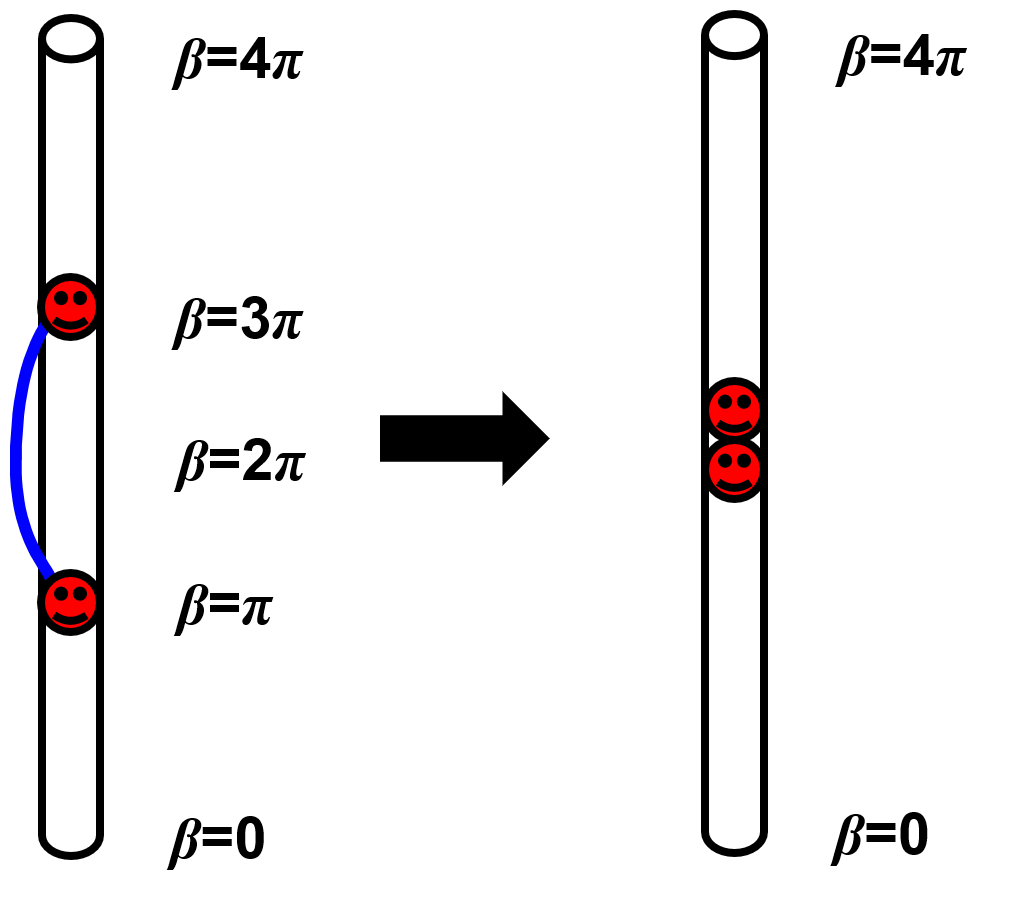}\\
(a) & (b)
\end{tabular}
\caption{Monopole(s) on a string attached by AB defect(s) 
in the deconfined phase. (a) a single monopole on a string  attached by an AB defect. 
Circles imply Wilson loops linking the string, 
which are trivial far from the monopole and exhibit 
an nontrivial AB phase around the monopole. 
(b) two monopoles on a string connected by an AB defect (left) 
and a monopole meson on a string (right).
\label{fig:monopole-on-string}
}
\end{figure}

To see this,
we consider a Wilson loop for the doublet 
encircling the string. 
We consider the doublet fields have a VEV in the second component 
and an Alice string has a winding in the upper-right component, 
as the first line Eq.~(\ref{eq:no-AB-defect}) with the upper sign.
Then, the Wilson loop 
is trivial for $\beta = 0,2\pi$ at $z = \pm \infty$.
However, the Wilson loop encircling around the monopole ($\beta =\pi$) 
is inevitably nontrivial, 
resulting in an AB defect (string) attached to the monopole 
as in Fig.~\ref{fig:monopole-on-string} (a).
In this sense, this can be regarded as a defect of the Wilson loop. 
The $U(1)$ modulus $\beta$ is a sine-Gordon-like soliton along the string.
If the AB defect is heavy enough compared with the tension of the Alice string, 
it will pull the string with constituting a $Y$-shape junction.

Next let us consider that the $U(1)$ modulus  $\beta$ of the Alice string winds twice along the string from $\beta =0$ at $z \to - \infty$ to $\beta =2\pi$ at 
$z \to + \infty$. 
Then, there exist two monopoles, and each of them is attached by  
an AB defect string as in Fig.~\ref{fig:monopole-on-string} (b) (left). 
Energetically, it is naturally to consider that these AB defects are connected, and the two monopoles are confined to be a monopole meson 
as in Fig.~\ref{fig:monopole-on-string} (b) (right).
Thus, the monopole confinement occurs on the Alice string.

\begin{figure}[!htb]
\centering
\hspace{-2cm}
\includegraphics[totalheight=4cm]{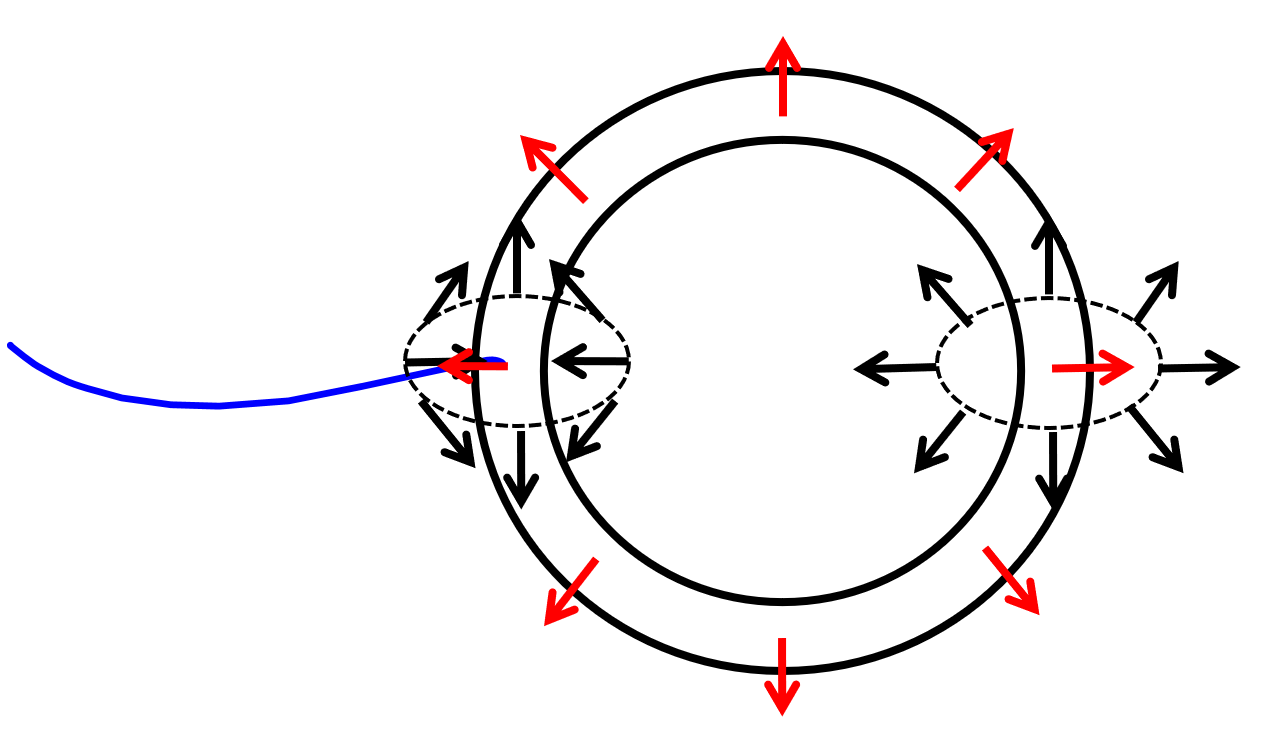}\\
(a)\\
\includegraphics[totalheight=4cm]{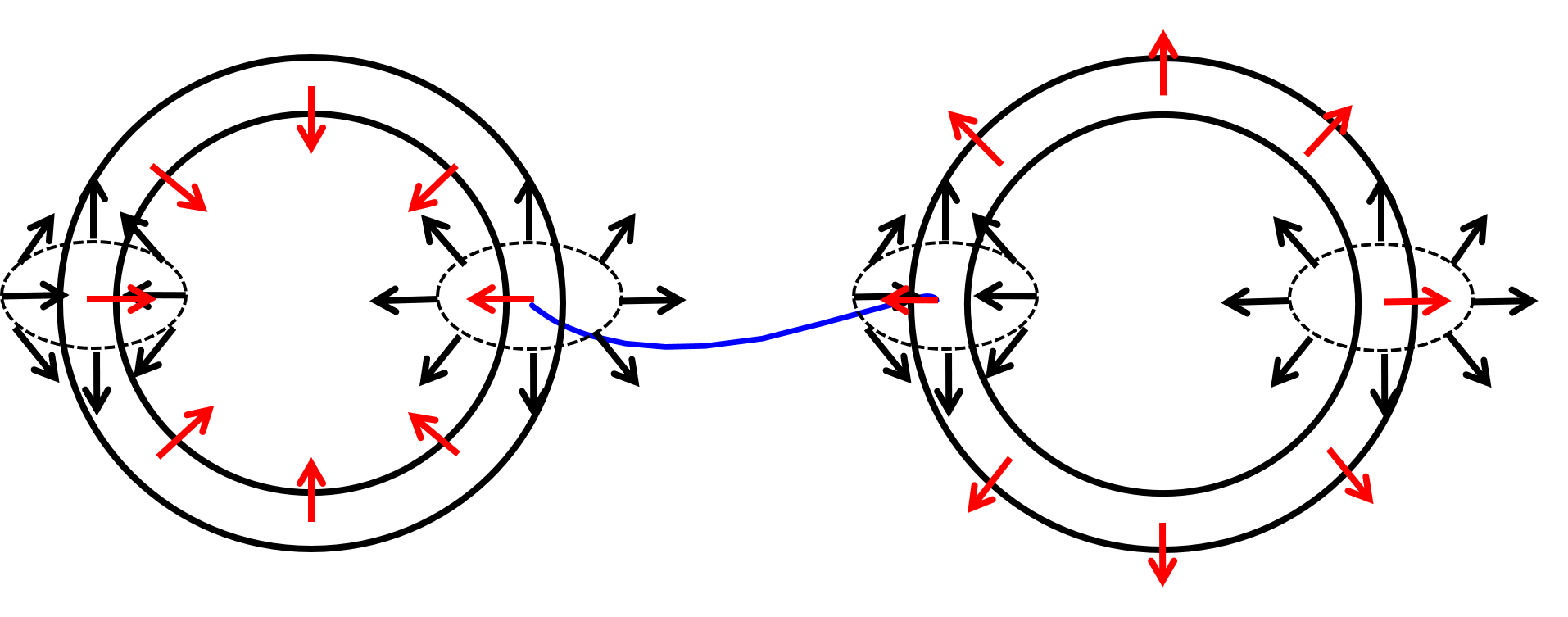}\\
(b)\\
\includegraphics[totalheight=3.2cm]{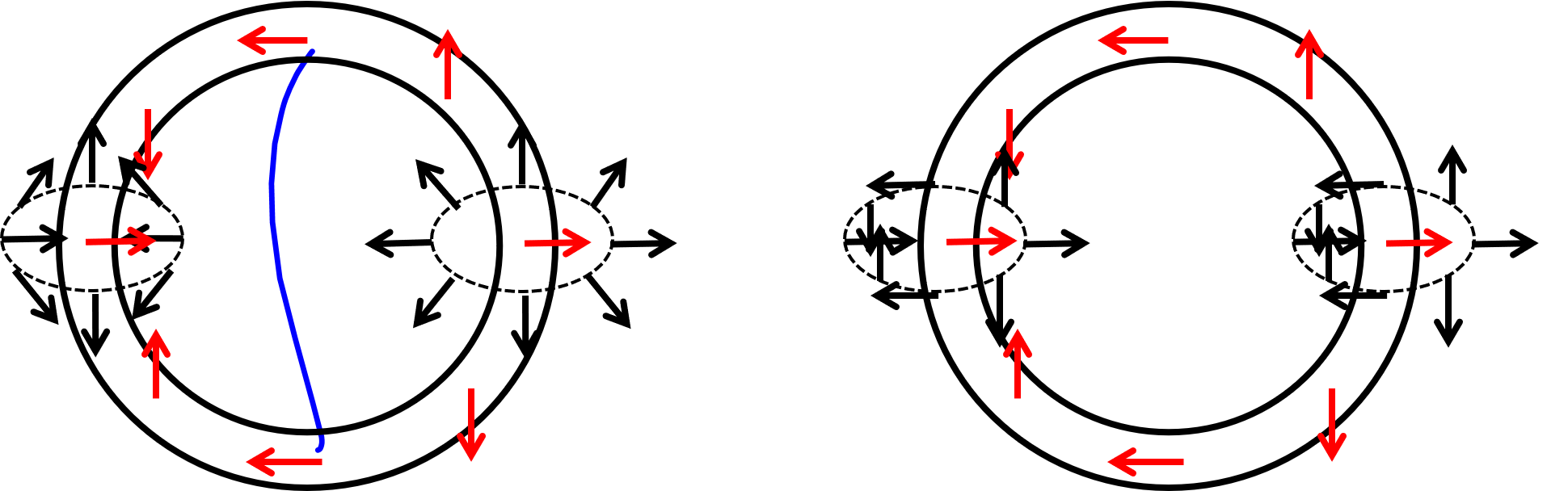}\\ 
(c)  \hspace{5cm} (d)
\caption{Confined monopoles in the deconfined phase. (a) A monopole of the charge one is attached by an AB defect. (b) Two monopoles of the charge one are connected by an AB defect.
(c) A doubly-twisted Alice ring as 
a monopole of the charge two.
(d) A singly twisted double-winding Alice ring as 
a monopole of the charge two.
\label{fig:confined-monopoles}
}
\end{figure}

\bigskip
Now, we are ready to discuss an isolated monopole.
An isolated monopole can be obtained as a twisted Alice ring 
as discussed in Sec.~\ref{sec:monopole}, 
which can be done by gluing the two ends of an Alice string 
in Fig.~\ref{fig:monopole-on-string}.
Then, as shown in Fig.~\ref{fig:confined-monopoles}(a), 
an AB defect is attached on the monopole point on the Alice ring. 
The $U(1)$ modulus also becomes a sine-Gordon-like soliton on a ring. 
This monopole must be confined.

As the case of Alice strings, 
it can be confined with an anti-monopole.
In that case, they are unstable and annihilate in pair.
On the other hand, a monopole can be confined with another monopole 
as in Fig.~\ref{fig:confined-monopoles}(b).
This configuration has the monopole charge two in total.
There are two more possibilities of the monopole of the charge two. 
One is 
a double winding of the  $U(1)$ modulus $\beta$ along the Alice ring in Fig.~\ref{fig:confined-monopoles}(c), 
and the other is a singly twisted $U(1)$ modulus $\beta$ 
along a ring of a doubly-wound Alice string  
as in Fig.~\ref{fig:confined-monopoles}(d).
Interestingly, the latter is not associated with any AB defect, 
since doubly-wound Alice strings do not have AB effects.
It is a highly nontrivial question 
depending on parameter choices 
which configurations 
among Fig.~\ref{fig:confined-monopoles}(b), (c) and (d) is the most stable.

In summary, monopoles of a single (or odd) monopole charge is confined, 
and monopoles of an even monopole charge are not confined and are allowed to exist.

\begin{figure}[!htb]
\centering
\begin{tabular}{cc}
\includegraphics[totalheight=5cm]{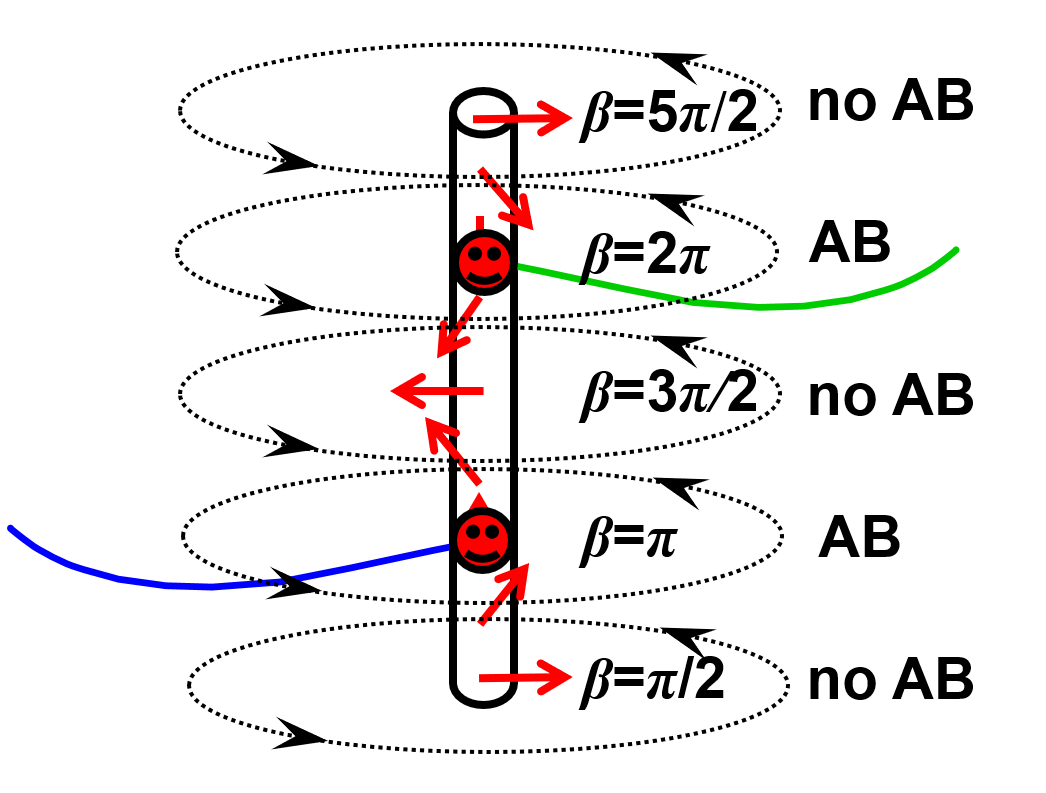} & \quad\quad\quad
\includegraphics[totalheight=5cm]{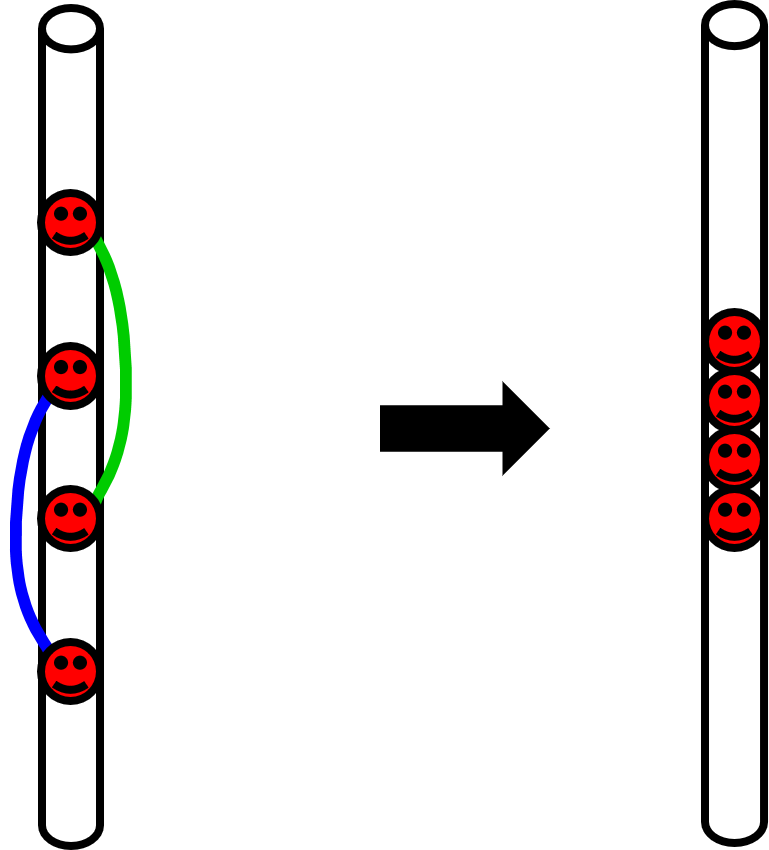}\\
(a) & (b)
\end{tabular}
\caption{Monopoles on a string in the confined phase. (a) a single monopole on a string is split into two half monopoles, each of which is attached by a different AB defect. 
(b) two monopoles or four half monopoles on a string 
connected by the two different AB defects (left), 
and a monopole meson on a string (right).
\label{fig:monopole-on-string-B}
}
\end{figure}

\bigskip
Let us discuss how monopoles look like in the deconfined phase.
In this case, AB defects appear around $\beta=0$ and $\beta=\pi$. 
Therefore, the AB defects disappear around $\beta = \pm \pi/2$. 
Monopoles on a straight string become like Fig.~\ref{fig:monopole-on-string-B}(a).
If we twist $\beta$ from once there appear two monopoles of the half charge,
each of which is attached by two different AB defects corresponding to two 
components of the doublet.
Therefore, in principle, only a half-monopole is possible to exist alone on a string.
The effective theory for $\beta$ would be double sine-Gordon-like.

If we consider monopoles of charge two on a string, 
they are confined as in Fig.~\ref{fig:monopole-on-string-B}(b).

\begin{figure}[!htb]
\centering
\hspace{-2cm}
\includegraphics[totalheight=4cm]{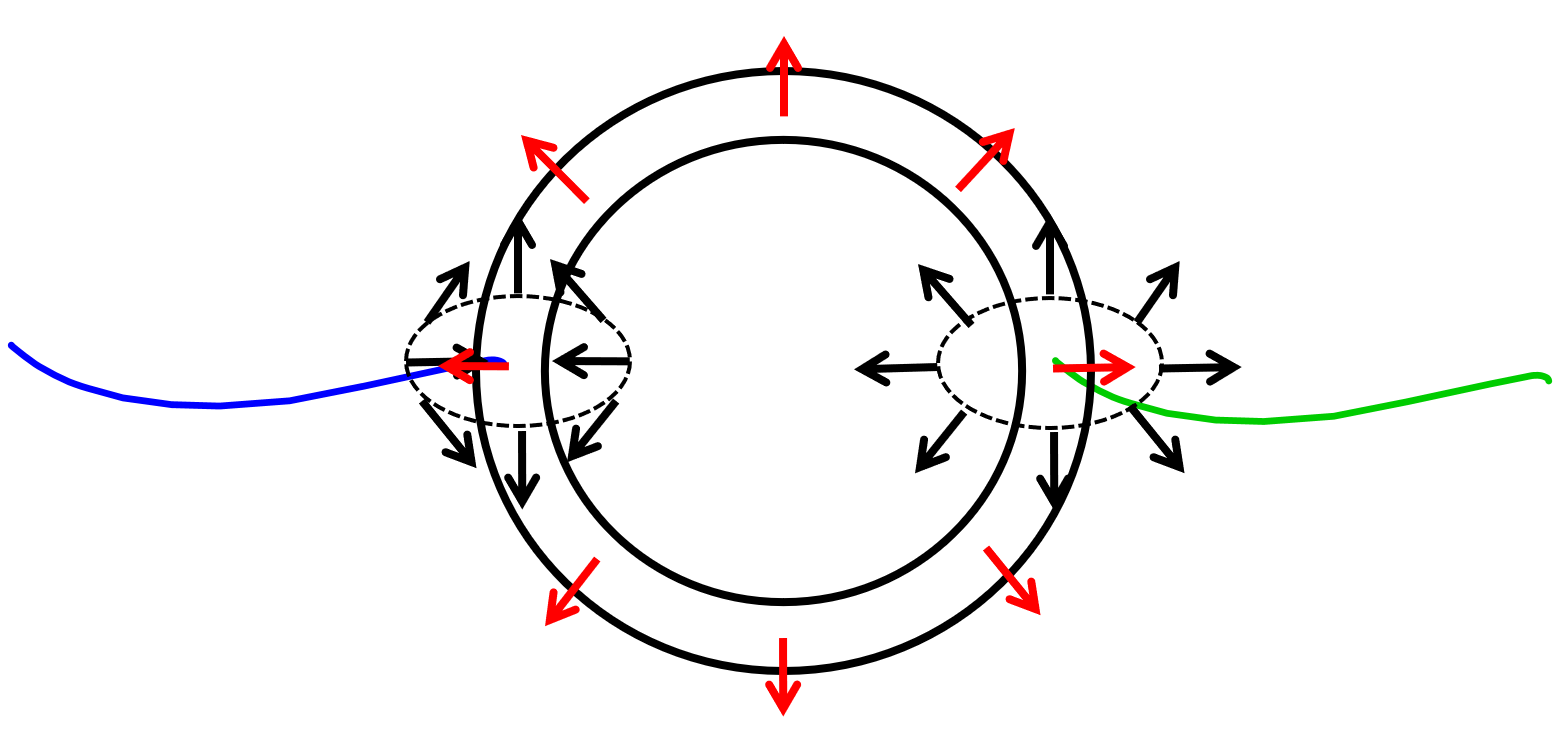}\\
(a)\\
\includegraphics[totalheight=4cm]{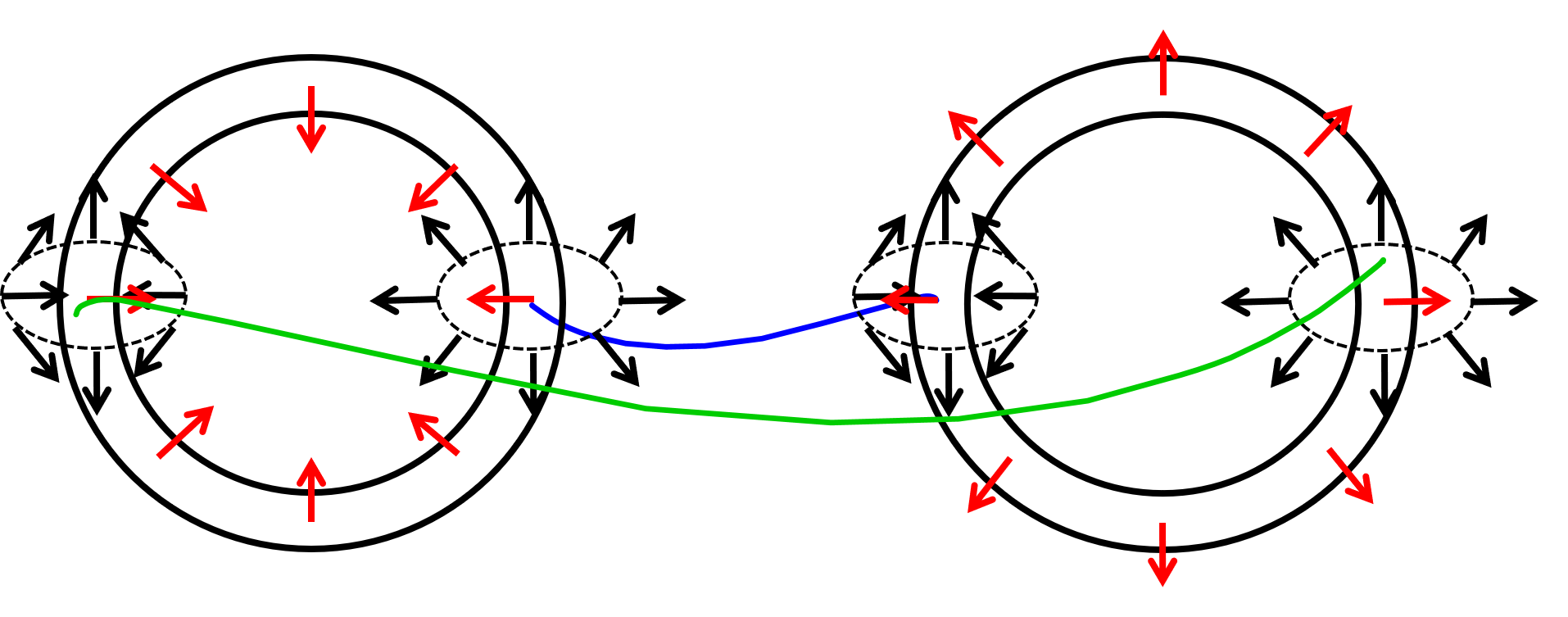}\\
(b)\\
\includegraphics[totalheight=3.2cm]{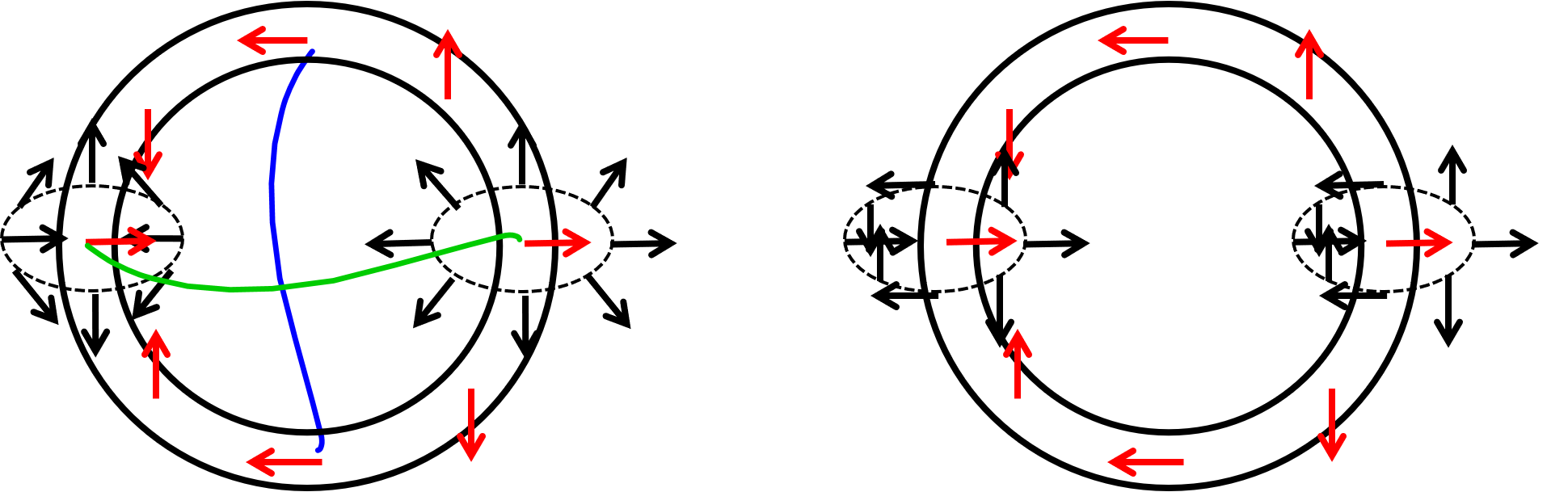}\\ 
(c)  \hspace{5cm} (d)
\caption{Confined monopoles in the confined phase. (a) A monopole of the charge one is attached by an AB defect. (b) Two monopoles of the charge one are connected by an AB defect.
(c) A doubly-twisted Alice ring as 
a monopole of the charge two.
(d) A singly twisted double-winding Alice ring as 
a monopole of the charge two.
\label{fig:confined-monopoles-B}
}
\end{figure}

On the other hand, isolated monopoles can be constructed as twisted 
Alice rings as before.
See Fig.~\ref{fig:confined-monopoles-B}.
A single monopole is attached by two AB defects as in (a). 
This must be confined as either (b) or (c):
two single monopoles are connected by two AB defects as in (b), 
and a doubly-twisted Alice ring have two AB defects connecting two sets of 
antipodal points 
of the ring as in (c).

\section{Summary and discussion}\label{sec:summary}

We have studied  
an $SU(2) \times U(1)$ gauge theory with charged triplet complex 
scalar fields admitting Alice strings and Alice monopoles, 
and introduced the charged doublet scalar fields 
receiving nontrivial AB phases around the Alice string. 
An ANO string and a ${\mathbb Z}_2$ string 
carry unit $U(1)$ magnetic flux and a half $SU(2)$ magnetic flux, respectively. 
First, we have studied the vacua in the presence of the doublet VEVs.
The unbroken symmetry is ${\mathbb Z}_4$ if one of VEVs is zero  (the deconfined phase), 
and it is ${\mathbb Z}_2$  if both VEVs are nonzero (the confined phase).
The Alice string carries a half $U(1)$ magnetic flux and 
$1/4$ $SU(2)$ magnetic flux taking a value in a linear combination of 
$\sigma^2$ and $\sigma^3$ 
(when the VEV of the triplet is $\sigma^1$).  
In the absence of a doublet VEV,
the Alice string is deconfined in the sense that 
the color flux can be seen by AB phases of some fields (like our doublet fields) encircling the string. 
When the doublet field develops VEVs in the presence of a single Alice string,
the bulk-soliton moduli locking occurs; 
the vacuum moduli are locked with the $U(1)$ modulus of the Alice string.  
The deconfined phase is realized to eliminate a domain wall called an AB defect 
responsible for single-valuedness of the doublet fields.
On the other hand, in the confined phase, 
the Alice string is inevitably attached by an AB defect.
Consequently, it is confined with either 
an anti-Alice string or the same type of an Alice string.
In the former case, the pair annihilates. 
In the latter case, 
the pair forms a stable doubly-wound Alice string 
having a color magnetic flux inside the core, 
but the color cannot be seen at large distance by AB phases 
and so the color confinement occurs.
We have referred these phases without and with doublet VEVs as the deconfined and confined phases, respectively. 
The results are also consistent with the vacuum manifold 
in Eq.~(\ref{eq:OPS}) 
admitting 1/4 quantized $SU(2)$ fluxes
for the deconfined phase,  
and the vacuum manifold in the confined phase in Eq.~(\ref{eq:G/K}) 
admitting only 1/2 quantized $SU(2)$ fluxes 
for the confined phase. 
In the scheme that the doublet field develops VEVs at high energy, 
the vacuum manifolds 
in Eq.~(\ref{eq:opposite-vacuum-mfd})
for the second symmetry breakings by 
the triplet field at low energy imply that the deconfined and confined phases allow 
1/4 and 1/2 quantized $W_3$ magnetic fluxes, 
which are a single Alice string and doubly wound Alice string, respectively, implying the confinement in the confined phase.
We also have shown that the ANO string and ${\mathbb Z}_2$ string decay into two Alice strings once the doublet field develops VEVs.
Furthermore, a monopole in this theory can be constructed as a closed Alice string 
with the $U(1)$ modulus twisted once, and 
we have found that
when the doublet field develops VEVs, monopoles are confined to mesons 
of the monopole charge two.

When a charged particle encircles an Alice string 
in the absence of the doublet VEVs,  
its electric charge is flipped.
This is an example of topological obstructions.
In addition to this, when a monopole encircles an Alice string, 
a monopole charge is also flipped, which is an example of topological influence 
(see Ref.~\cite{Kobayashi:2011xb} for a global analogue of this phenomenon). 
When the doublet field develops VEVs, 
Alice strings are confined to doubly-wound Alice strings 
and monopoles are confined to monopole mesons 
as discussed in this paper.
In this case, it seems that the topological obstructions and topological influence disappear.
This suggests that deconfinement and 
confinement phases may be characterized by the presence and absence, respectively  
of the topological obstruction and topological influence.

Several discussions are addressed here.

   If we turn off the $U(1)$ gauge coupling, $e=0$, 
   and regard this global $U(1)$ symmetry as 
   the Peccei-Quinn symmetry for axions,
   our model becomes an axion dark matter model 
     in Refs.~\cite{Sato:2018nqy,Chatterjee:2019rch}.
  After chiral symmetry breaking, 
  an axion string is attached by two axion domain walls 
  and the domain wall number is two in the model.
  When the domain wall number is more than one,
  usually there is a domain wall problem 
(if domain walls cannot decay and 
dominate Universe) \cite{Kawasaki:2013ae}.
In our case, one axion string attached by two domain walls can decay 
into two Alice strings each of which is attached by only one domain wall, thereby free from the domain wall problem. 
In Ref.~\cite{Sato:2018nqy}, the doublet scalar field was also 
considered. This doublet field receives an AB phase as discussed in the present paper. If the doublet develops a VEV, one axion string (corresponding to ANO string 
in this paper) can decay into two Alice strings,
thereby a domain wall problem does not occur even before chiral symmetry breaking.

The third homotopy group for the first symmetry breaking by the triplet  in Eq.~(\ref{eq:homotopy-G/H}) implies the existence of gauged Hopfions 
 (for a global analogues of the same breaking, see Ref.~\cite{Kawaguchi:2008xi}).
The second homotopy group for the second symmetry breaking by the doublet  in Eq.~(\ref{eq:homotopy-doublet}) implies that 
there are another type of vortices in the both deconfined and confined phases.
   Studying these topological objects remains as a future problem.

   Fermions receiving non-trivial AB phases can contribute to AB defects 
   if they form pair condensations. 
   In fact, such an example can be found in dense QCD 
   \cite{Fujimoto:2020dsa}, in which 
   an $SU(3)$ generalization of our model was discussed.

\section*{Acknowledgment}
The author thanks Chandrasekhar Chatterjee for discussion in the previous works and at the early stage of this work.
This work is  supported in part by 
JSPS Grant-in-Aid for Scientific Research 
(KAKENHI Grant 
No.~18H01217).

\begin{appendix}

\section{Complex symmetric tensor}\label{sec:app0}
Here, we summarize an equivalence
between a complex adjoint scalar field 
and a complex symmetric tensor scalar field.

A complex adjoint scalar field 
considered in this paper 
\begin{eqnarray}
 \Phi = \sum_{\alpha=1}^3 \Phi^\alpha \sigma^\alpha
\end{eqnarray}
transforms as
\begin{eqnarray}
 \Phi \to \Phi' = e^{i\alpha} g \Phi g^{-1}, \quad 
 (e^{i\alpha},g) \in U(1)_Y \times SU(2)_W.
\end{eqnarray}
On the other hand, 
a complex $2\times 2$  symmetric tensor scalar field
\begin{eqnarray}
 M = M_0 {\bf 1}_2 + M_1 \sigma^1 + M_3 \sigma^3,
\end{eqnarray}
transforms as
\begin{eqnarray}
 M \to M' = e^{i\alpha} g M g^T,
   \quad 
 (e^{i\alpha},g) \in U(1)_Y \times SU(2)_W.
\end{eqnarray}
Both fields have three complex scalar degrees of freedom.

From the relations  
\begin{eqnarray}
&& (+i\sigma^2) ({\bf 1}_2, \vec{\sigma}^T ) (-i\sigma^2) = ({\bf 1}_2, -\vec{\sigma}), 
 \nonumber\\ 
&& (i\sigma^2) g^T (-i\sigma^2) = g^{-1}, \quad g \in SU(2),
\end{eqnarray}
we can easily see that $ M (-i\sigma^2) $ in fact transforms 
as an adjoint representation,
thus implying the relation between them, 
\begin{eqnarray}
 && \Phi = M (-i\sigma^2) = -i M_0 \sigma^2 + M_1 \sigma^3 - M_3 \sigma^1,\\
&& (\Phi^1, \Phi^2, \Phi^3) = (-M_3, -i M_0, M_1).
\end{eqnarray}
From the coefficient $i$ in the second component, one can understand that 
this relation holds only between complex scalar fields but not 
between real scalar fields.

The vacuum $\Phi = \Phi^1 \sigma^1$ in Eq.~(\ref{eq:gauge1}) 
considered in this paper 
corresponds to $M = M_3 \sigma^3$.
The unbroken subgroup of the $M = \xi \sigma^3$ is 
$H = ({\mathbb Z}_4)_{Y+W_{2,3}} \ltimes SO(2)_{W_1}$ 
with the $SO(2)$ part generated by $\sigma_1$,
which of course does not depend on 
a choice of representation $\Phi$ or $M$, equivalent to each other.

In terms of the symmetric tensor $M$, 
the ANO and ${\mathbb Z}_2$ strings can be written as
\begin{eqnarray}
&&  M (r, \theta)  
\sim 
 \xi 
 \left(
\begin{array}{ccc}
  e^{ i \theta}  &  0     \\
   0 & -e^{ i \theta}     
\end{array}
\right) 
=
 \xi 
  e^{ i \theta} \sigma^3
, \\
&& M (r, \theta)  
\sim  
 \xi 
 \left(
\begin{array}{ccc}
  e^{ i \theta}   &  0     \\
   0 & -e^{ - i \theta}    
\end{array}
\right) 
= \xi e^{i\frac{\theta}{2}\sigma^3} \sigma^3 e^{i\frac{\theta}{2}\sigma^3},
\end{eqnarray}
with 
the gauge field configurations in 
Eqs.~(\ref{eq:ANO}) and 
(\ref{eq:Alice-asymptotic-Z2}), respectively. 

The Alice strings corresponding to Eq.~(\ref{eq:Alice-asymptotic}) and (\ref{eq:Alice-asymptotic2}) can be written as
\begin{eqnarray}
&& M(r, \theta)  
\sim 
 \xi \left(
\begin{array}{cc}
  e^{ i \theta} & 0 \\
  0 & 1       
\end{array}
\right) = 
\xi e^{i\frac{\theta}{2}}  \left(
\begin{array}{ccc}
       e^{+i\frac{\theta}{2}}  &  0      \\ 
        0 & - e^{-i\frac{\theta}{2}}  
\end{array}
\right) 
= \xi e^{i\frac{\theta}{2}} e^{+i\frac{\theta}{4}\sigma^3} \sigma^3 e^{+i\frac{\theta}{4}\sigma^3}, 
  \label{eq:Alice-asymptotic-M}  \\
&& M(r, \theta)  
\sim  \xi \left(\begin{array}{cc}
  1 &  0 \\
  0 & - e^{ i \theta}       
\end{array}
\right) 
= \xi e^{i\frac{\theta}{2}} \left(
\begin{array}{ccc}
       e^{-i\frac{\theta}{2}}  &  0      \\ 
        0 & - e^{+i\frac{\theta}{2}}  
\end{array}
\right) 
= \xi e^{i\frac{\theta}{2}} e^{-i\frac{\theta}{4}\sigma^3} \sigma^3 e^{-i\frac{\theta}{4}\sigma^3}, 
  \label{eq:Alice-asymptotic2-M}
\end{eqnarray}
with the same gauge field configurations in 
Eq.~(\ref{eq:Alice-asymptotic}) and (\ref{eq:Alice-asymptotic2}), respectively, 
while their anti-strings in Eqs.~(\ref{eq:Alice-asymptotic-anti}) and 
(\ref{eq:Alice-asymptotic2-anti})
can be written as
\begin{eqnarray}
&& M(r, \theta)  
\sim 
\xi \left(\begin{array}{cc}
  1 &  0 \\
  0 & - e^{- i \theta}       
\end{array}
\right) 
= 
\xi e^{-i\frac{\theta}{2}}  \left(
\begin{array}{ccc}
       e^{+i\frac{\theta}{2}}  &  0      \\ 
        0 & - e^{-i\frac{\theta}{2}}  
\end{array}
\right) 
= \xi e^{- i\frac{\theta}{2}} e^{+i\frac{\theta}{4}\sigma^3} \sigma^3 e^{+i\frac{\theta}{4}\sigma^3},\\
&& M(r, \theta)  
\sim
 \xi \left(
\begin{array}{cc}
  e^{- i \theta} & 0 \\
  0 & 1       
\end{array}
\right)  
= \xi e^{-i\frac{\theta}{2}} \left(
\begin{array}{ccc}
       e^{-i\frac{\theta}{2}}  &  0      \\ 
        0 & - e^{+i\frac{\theta}{2}}  
\end{array}
\right) 
= \xi e^{- i\frac{\theta}{2}} e^{-i\frac{\theta}{4}\sigma^3} \sigma^3 e^{-i\frac{\theta}{4}\sigma^3}, 
\end{eqnarray}
respectively, with the same gauge field configurations.

The doubly-wound Alice strings corresponding to Eq.~(\ref{eq:Alice-asymptotic-double}) and (\ref{eq:Alice-asymptotic-double2}) can be written as
\begin{eqnarray}
&& M(r, \theta)  
\sim 
 \xi \left(
\begin{array}{cc}
  e^{ 2i \theta} & 0 \\
  0 & 1       
\end{array}
\right) = 
\xi e^{i \theta}  \left(
\begin{array}{ccc}
       e^{+i\theta}  &  0      \\ 
        0 & - e^{-i\theta}  
\end{array}
\right) 
= \xi e^{i\theta} e^{+i\frac{\theta}{2}\sigma^3} \sigma^3 e^{+i\frac{\theta}{2}\sigma^3}, 
  \label{eq:Alice-asymptotic-double-M}  \\
&& M(r, \theta)  
\sim  \xi \left(\begin{array}{cc}
  1 &  0 \\
  0 & - e^{ 2i \theta}       
\end{array}
\right) 
= \xi e^{i\theta} \left(
\begin{array}{ccc}
       e^{-i\theta}  &  0      \\ 
        0 & - e^{+i\theta}  
\end{array}
\right) 
= \xi e^{i\theta} e^{-i\frac{\theta}{2}\sigma^3} \sigma^3 e^{-i\frac{\theta}{2}\sigma^3}, 
  \label{eq:Alice-asymptotic-double2-M}
\end{eqnarray}
with the same gauge field configurations in 
Eq.~(\ref{eq:Alice-asymptotic-double}) and (\ref{eq:Alice-asymptotic-double2}), respectively.

\section{Alice string configurations}
\label{sec:app}

Here, we briefly summarize BPS Alice strings 
  \cite{Chatterjee:2017jsi,Chatterjee:2017hya}
 in the absence of the doublet field.  
setting all the scalar couplings to be zero except $\lambda_g$ and $\lambda_e$. 
The static Hamiltonian density is given by
\begin{eqnarray}
 \label{alice_Hamiltonian}
&&{H} = \int d^3x \left[\frac{1}{2} \Tr F_{ij} F^{ij} + \frac{1}{4} f_{ij}f^{ij} + \Tr | D_i\Phi|^2  
+  \frac{\lambda_g}{4} \Tr[\Phi,\Phi^\dagger]^2
+  \frac{\lambda_e}{2}\left(\Tr \Phi\Phi^\dagger  - 2 \xi^2\right)^2\right]. \nonumber\\
\end{eqnarray}
At the critical couplings 
$ \lambda_e = e_{}^2$ and $\lambda_g = g^2$, 
the Bogomol'nyi completion of the tension 
can be performed as 
\begin{eqnarray}
\label{tension}
\mathcal{T} & =& \int d^2x \left[\Tr\left[ F_{12} \pm \frac{g}{2}[\Phi,\Phi^\dagger]\right]^2 
+ \Tr|\D_{\pm} \Phi|^2 
+ \half\left[  f_{12} \pm e_{} \left(\Tr \Phi\Phi^\dagger - 2 \xi^2\right)\right]^2
\pm 2 e_{} f_{12} \xi^2 \right.\Big{]} \nonumber\\ 
 &&\ge 2 e_{}  \xi^2 \left|\int d^2x\; f_{12}\right| = 2 \pi  \xi^2 ,\;
\end{eqnarray}
with $\D_{\pm} \equiv \frac{D_1 \pm i D_2}{2}$.  
The equality is achieved if 
the BPS equations are satisfied: 
\begin{eqnarray}
&& F_{12} \pm \frac{g}{2}[\Phi,\Phi^\dagger]=0, \nonumber\\
&& \D_{\pm} \Phi =0, \nonumber\\
&& f_{12} \pm e_{} \left(\Tr \Phi\Phi^\dagger - 2 \xi^2\right)=0.\label{eq:BPS}
\end{eqnarray}

Alice string solution 
can be constructed by inserting profile functions 
$f_1(r), f_2(r), A(r)$ and $a(r)$ 
into the asymptotic form in Eq.~(\ref{eq:Alice-asymptotic}):
\begin{eqnarray}\label{phivortex}
&&\Phi(r, \theta) = 
\xi\left(
\begin{array}{ccc}
 0 & f_1(r) e^{i\theta}     \\
f_2(r)   & 0  \end{array}
\right), \nonumber\\
&& A_i(r, \varphi) = -\frac{1}{4g} \frac{\epsilon_{ij} x_j}{r^2}\sigma^3A(r), \quad 
 a_i (r, \theta) = - \frac{1}{2e_{}} \frac{\epsilon_{ij}x_j}{r^2}a(r), 
\end{eqnarray}
where $(r, \theta )$ are radial and angular coordinates 
of the two dimensional space, 
respectively. 
The profile functions
depend only on the radial coordinate 
and 
can be solved numerically 
with the boundary conditions 
$f_1(0) = f_2'(0) = 0,\quad f_1(\infty) = f_2(\infty) =1, A(0)= a(0) = 0, \quad A(\infty)= a(\infty) = 1$.\footnote{
These profile functions eventually satisfy the same equations with 
those for a non-Abelian vortex in $U(N)$ gauge theory 
coupled with $N$ Higgs scalar fields in the fundamental representation 
\cite{Hanany:2003hp,Auzzi:2003fs,
Eto:2004rz,
Eto:2005yh,Eto:2006cx,
Tong:2005un,Eto:2006pg,Shifman:2007ce,Shifman:2009zz}. 
}
The numerical solution for the BPS equation (\ref{eq:BPS})
can be found in Ref.~\cite{Chatterjee:2017jsi}.

\begin{figure}[htbp]
\centering
\begin{tabular}{ccc}
\includegraphics[totalheight=3.5cm]{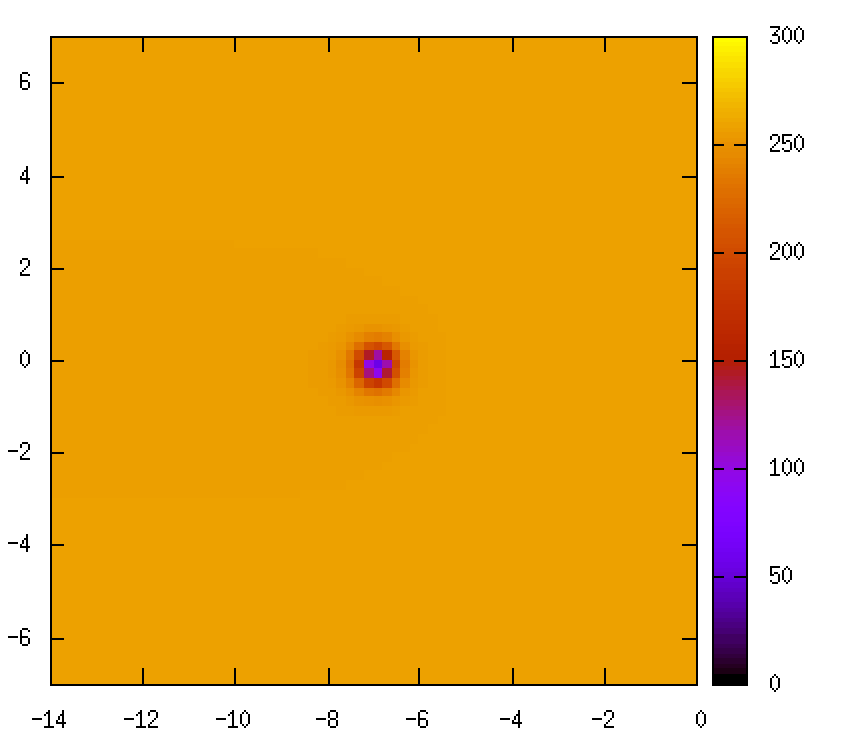} 
&
\includegraphics[totalheight=3.5cm]{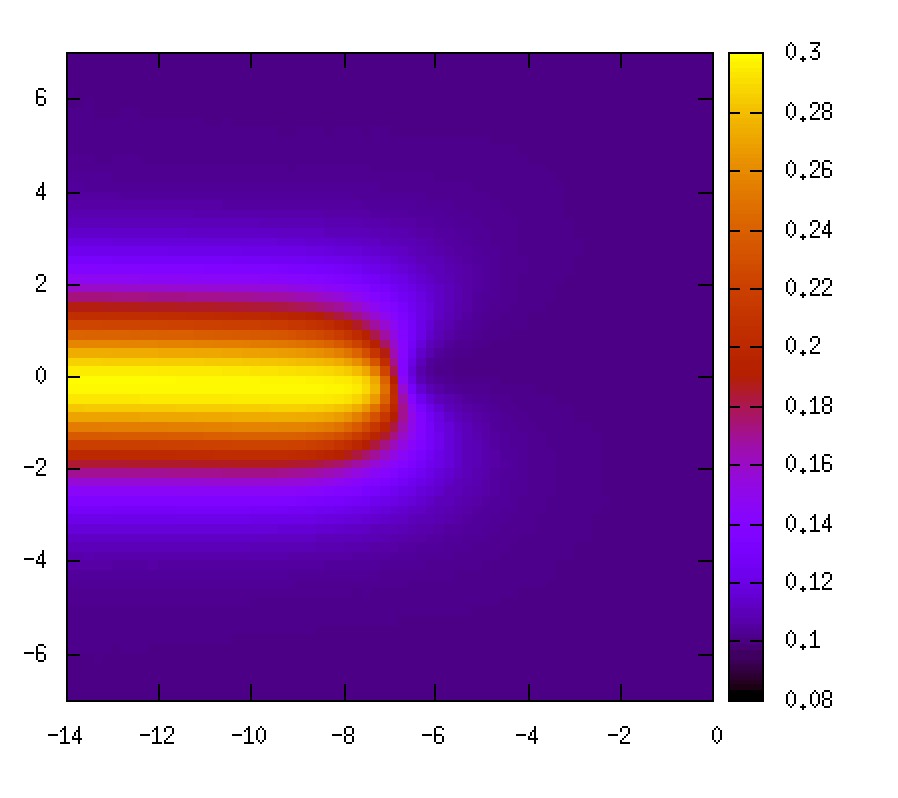} 
& 
\includegraphics[totalheight=3.5cm]{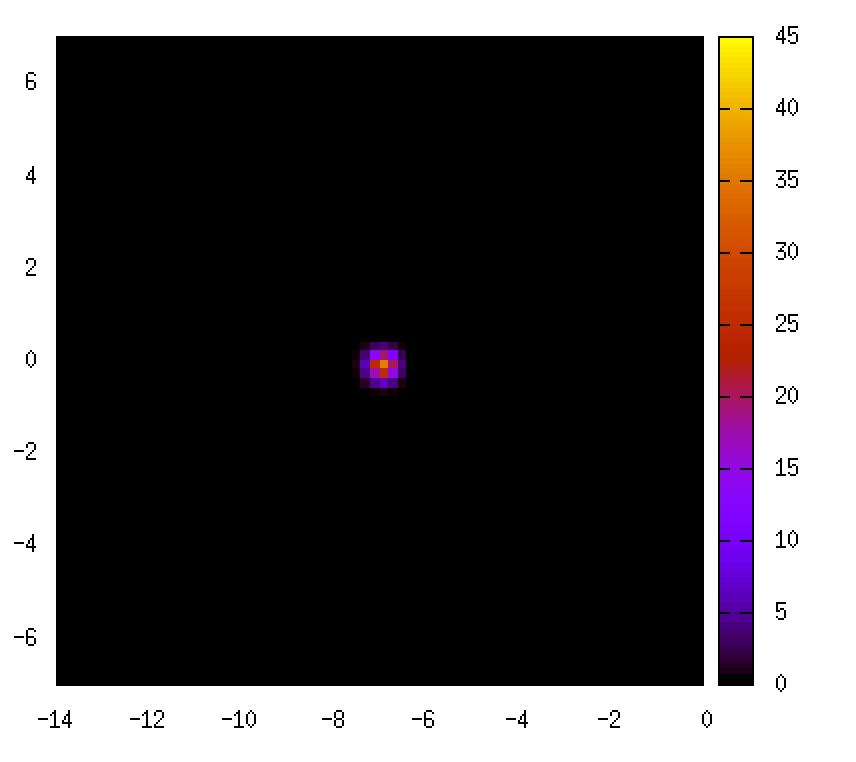}
 \\
(a) & (b) & (c) 
\end{tabular}
\caption{A full 2D numerical simulation for an Alice string confined by a soliton (figures taken from \cite{Chatterjee:2019zwx}). 
(a) The gauge invariant 
$4 \Tr\Phi^2 \Tr{\Phi^\dagger}^2\sim \xi^4 f(r)^2g(r)^2$ vanishing at the center of the Alice string.
 (b) The interaction potential $V_{\rm int}^{(1)} + \frac{2\xi m^2}{\lambda_\psi}$, 
 in which the domain wall is clearly visible.
  (c) The flux squared $\Tr F_{12}^2 =\sum_{a=1}^3{F_{12}^a}^2$.
 The parameter choices are  $\xi = 0.1,   e_{} = 0.5,  g = 1,
     \mu^{(1)}  = 0.1 , \lambda_\psi= 1, m =  0.1, \lambda_e= 2, \lambda_g=1 $. 
     }
\label{DW_Alice1}
\end{figure}

Once the doublet field develops VEVs, 
only 2D full numerical solution of an Alice string attached 
by an AB defect is available 
\cite{Chatterjee:2019zwx}, see Fig.~\ref{DW_Alice1}.

 \end{appendix}


\end{document}